\nonstopmode
\documentclass[useAMS,usenatbib]{mn2e}
\usepackage{graphicx}

\title[High Galactic latitude polarized emission at 1.4 GHz]
      {High Galactic latitude polarized emission at 1.4 GHz and implications
for cosmic microwave background observations}
\author[E. Carretti, et al.]
{E. Carretti$^{1}$\thanks{E-mail:
carretti@bo.iasf.cnr.it},
G. Bernardi$^{1}$, R.J. Sault$^{2}$,
S. Cortiglioni$^{1}$ and S. Poppi$^{3}$\\
$^{1}$CNR--IASF Bologna, Via Gobetti 101, Bologna, I-40129, Italy\\
$^{2}$CSIRO--ATNF, P.O. Box 76, Epping, NSW 1710, Australia\\
$^{3}$CNR--IRA Bologna, Via Gobetti 101, Bologna, I-40129, Italy}
\begin{document}

\date{Accepted xx xx xx. Received yy yy yy; in original form zz zz zz}

\pagerange{\pageref{firstpage}--\pageref{lastpage}} \pubyear{2004}

\maketitle

\label{firstpage}

\begin{abstract}
We analyse the polarized emission at 1.4~GHz in a $3^\circ\times 3^\circ$
area at high Galactic latitude ($b \sim -40^\circ$).
The region, centred in ($\alpha = 5^{\rm h}$,~$\delta = -49^\circ$),
was observed with the Australia Telescope Compact Array
radio-interferometer,
whose 3--30~arcmin angular sensitivity range allows the study of
scales appropriate for Cosmic Microwave Background Polarization (CMBP)
investigations.
The angular behavior of the diffuse emission is analysed through the
$E$- and $B$-mode angular power spectra. These follow
a power law $C^X_\ell \propto \ell^{\beta_X}$ with slopes
$\beta_E = -1.97 \pm 0.08$ and $\beta_B = -1.98 \pm 0.07$.
The emission is found to be about a factor 25 fainter
than in Galactic plane regions. The
comparison of the power spectra with other surveys indicates that
this area is intermediate between strong
and negligible Faraday rotation effects.
A similar conclusion can be reached by analysing both the frequency and
Galactic latitude behaviors of the diffuse Galactic emission of the
408-1411~MHz Leiden survey data.
We present an analysis of the Faraday rotation effects on the polarized
power spectra, and find that
the observed power spectra can be enhanced by a transfer of power
from large to small angular scales. 
The extrapolation of the spectra to 32 and 90~GHz of the CMB window 
suggests that Galactic synchrotron emission leaves the CMBP $E$-mode
uncontaminated at 32~GHz. The level of the contamination at 90~GHz is 
expected to be more than 4 orders of magnitude
below the CMBP spectrum.
Extrapolating to the relevant angular scales,
this region also appears adequate for investigation of the
CMBP $B$-modes for models with tensor-to-scalar fluctuation 
power ratio $T/S > 0.01$.
We also identify polarized point sources in the field,
providing a 9 object list which is complete down to the polarized flux limit
of $S^p_{\rm lim} = 2$~mJy.
\end{abstract}


\section{Introduction}
Polarized Galactic synchrotron emission is one of the most
important foregrounds in measuring the
Cosmic Microwave Background Polarization (CMBP) up to about 100~GHz.
Above that frequency dust emission becomes the dominant contaminant.
Moreover, as the fractional polarization of the synchrotron emission can
be 30\% and as CMBP is just few percent of the CMB anisotropy,
the degree of contamination is expected to
be worse than in temperature anisotropy measurements.

The study of this foreground
is thus crucial for CMBP experiments.
It will allow an estimate of the
contamination level, 
and will aid developing cleaning procedures to remove its
contribution from the cosmic
signal (e.g. see \citealt{tegmark00} and references therein).

The CMBP emission peaks on sub-degree angular scales in the 5--30~arcmin range
(e.g. see~\citealt*{zaldarriaga97b}). Consequently, observations of CMBP
can be carried on in small sky patches that are large enough to allow
good statistics ($5^\circ$--$10^\circ$ wide) but, at the same time,
small enough to minimize environmental systematics,
e.g. spillover from ground emission. The latter could swamp the detection
of the faint (few $\mu$K) cosmic signal.
Small regions, moreover, have the additional advantage of allowing
a part of the sky low in synchrotron
contamination to be selected.

The synchrotron emission can be best observed at low frequency, where
it dominates other diffuse components (dust, free-free,
and CMBP itself) and where the signal strength allows easier detection by
radiotelescopes.
To date, analysis of the Galactic synchrotron has been performed
by several surveys at frequencies up to 2.7~GHz.
These have been mainly concentrated on the Galactic plane, where the
signal is
greater.
The Southern Galactic Plane Survey (SGPS,~\citealt{gaensler01}) and the
Canadian Galactic Plane Survey (CGPS,~\citealt{taylor03})
are interferometric surveys whose main goal is to map almost all the
Galactic plane at 1.4~GHz down to a 1~arcmin resolution.
These surveys provide deep insight into
many effects typical of Galactic polarized emission (e.g., polarization
horizons, Faraday screens). However, because of their interferometric
origin, these surveys are sensitive to angular
scales no larger than the 30~arcmin of the telescope primary beam.

The largest angular scales are instead covered by a single dish
project:
the Effelsberg 1.4~GHz Medium Galactic Latitude Survey (EMLS, \citealt{uyaniker99}
and~\citealt{reich04}). This covers the Galactic plane up to medium
latitudes ($|b|<20^\circ$).
The approximately 10~arcmin resolution allows EMLS to overlap
the scales accessible by the two previous interferometric surveys,
and their combination will provide full information on the Galactic plane
down to 1~arcmin.

The Galactic plane has been also surveyed at higher frequency
by~\citet{duncan97} and ~\citet{duncan99}. They covered about a half
of the plane at 2.4 and 2.7~GHz, respectively, with the
latitude coverage extends to $|b| < 5^\circ$.

The mid Galactic latitudes has been partially surveyed at 350~MHz by
the Westerbork survey~\citep*{have03},
which mapped the polarized emission at longitudes
$l\sim 140^\circ$--170$^\circ$ and up to $b = 30^\circ$.
Because of their lower frequency, these data
are likely to be more affected by Faraday rotation effects.

The analysis of the Galactic plane region has provided
the first information on the angular behavior of
the Galactic synchrotron radiation (see~\citealt{br02}
and references therein). However, the optimal
locations for observations of the CMBP are at
high latitudes, where the emission is low.
High latitude regions are not as well studied, although more recent
surveys are starting to fill this gap, at least at 1.4~GHz.

Sparse observations at all Galactic latitudes
were made in 1960s and 1970s (e.g. \citealt{baker74,bs76},
and references therein). Among them, the 
observations presented by \citet{bs76} covered all Galactic latitudes
at 5 frequencies between 408 and 1411~MHz. For decades these 
have represented the largest data set of polarized measurements 
out of the Galactic plane.
Although undersampled and not suitable for a full investigation
of the synchrotron characteristics (especially when dealing
with the angular behavior) some analyses of the angular power spectra
have been carried out in the best sampled portions~\citep{br02}.
Moreover, as we will see in this work,
this data-set can provide useful information on the frequency behavior
of the large angular scale structure.

The whole of the Northern and Southern sky are being
mapped by \citet{wolleben04} and \citet*{testori04}, respectively, at 1.4~GHz.
These surveys are very precious, since they
will provide the first all-sky map of the polarized emission at 1.4~GHz down to
the degree scale.
They are sensitive to the large angular scale structure
of the polarized emission in interesting (for CMBP purposes)
low emission regions.
However, their angular resolution (larger than 30~arcmin)
and sensitivity (about 15~mK) do not allow
the analysis of the faint areas required by
sub-degree CMBP experiments, which require better resolution
(finer than 30~arcmin) and sensitivity
(rms signal about 10~mK, see \citealt{be03}).

All of the aforementioned surveys are at low frequencies, 
where the Faraday rotation effects
can still play a significant role. In fact, Faraday rotation
can introduce a randomization of polarization angles which
transfers power from large to small angular scales
by modifying the polarized emission pattern.
Although this transfer of power has been claimed to explain some
results at frequencies
up to 1.4~GHz (e.g ~\citealt{tu02}), it has not been
quantitatively studied. A detailed study is
needed to evaluate its effect before we can safely extrapolate up
to the mm-wave cosmological window.

The patch of sky near $\alpha = 5^{\rm h}$ and $\delta = -50^{\circ}$
has been identified as an interesting area for CMBP investigations:
it is at high Galactic latitude
($b \sim -40^\circ$) and, from an analysis of the Rhodes/HartRAO 2326-MHz radio
continuum survey \citep*{jonas98},
it is expected to have a synchrotron emission at the low level required
for CMBP studies~\citep{bernardi04,carretti02}.
In fact, this region was chosen for the
total intensity observation by the BOOMERanG experiment
and has been selected as target for BaR-SPOrt~\citep{cortiglioni03}
and BOOMERanG-B2K~\citep{masi02}.

This patch has been observed
in polarization at 1.4~GHz to a
sensitivity to allow detection of the polarized synchrotron
emission \citep{be03}. This represents the first detection of this emission
at high Galactic latitude at this frequency and
in the angular scale range useful
for CMBP analyses (5--30~arcmin).

These observations, along with an initial analysis of the
implications for CMBP studies, have been presented by \citet{be03}.
In the present paper we perform a more detailed analysis of
the polarized emission in this area and reach
firmer conclusions regarding CMBP implications.

\begin{figure*}
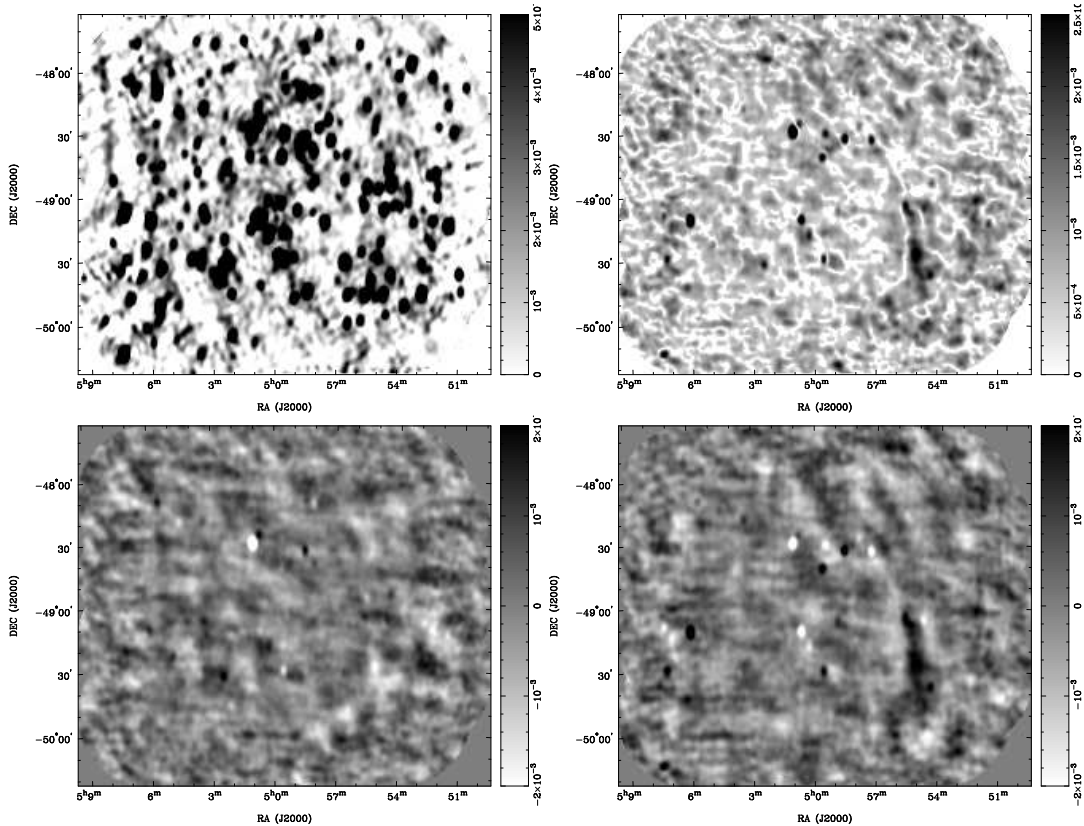

  \includegraphics[angle=-90, width=0.4\hsize]{fig1a.ps}
  \includegraphics[angle=-90, width=0.4\hsize]{fig1b.ps}
  \includegraphics[angle=-90, width=0.4\hsize]{fig1c.ps}
  \includegraphics[angle=-90, width=0.4\hsize]{fig1d.ps}
  \caption{From top-left, clockwise: $I$, $I^p$, $U$ and $Q$
	   maps at 1.4~GHz of the $3^\circ \times 3^\circ$ area centred
	   on $\alpha = 5^{\rm h}$, $\delta = -49^{\circ}$. Values are Jy~beam$^{-1}$.
	   Point sources have not been subtracted. (See~\citealt{be03} for
	   polarized maps cleaned from point sources.)}
  \label{iquFig}
\end{figure*}

We quantatively analyse the effects of the polarization angle
randomization introduced by Faraday screens,
finding that a power transfer
from large to small angular scales occurs. This results both 
in a steepening of the (angular) spectral index and 
an enhancement of the observed
emission on the angular scales our observations are sensitive to
(i.e. 3--30~arcmin). This implies that our measurements represent an upper
limit of the intrinsic polarized emission.

We present the angular power spectra of the $E$- and $B$-modes.
Comparison of these with other surveys suggests that the observed
patch is an intermediate state
between significant and negligible Faraday rotation effects.

Finally, we extrapolate the spectra up to the frequency range of CMB measurements
(30--90~GHz). This suggests that the Galactic synchrotron emission
is low enough in this area that it should not be an issue for both 
the $E$-mode (from 30~GHz) and the 
fainter $B$-mode (provided a tensor-to-scalar power ratio $T/S > 0.01$).

The paper is organized as follows. In Section~\ref{obsSec} we present a
quick summary of the observations. In Sections~\ref{pointSec} and~\ref{rmSec}
we analyse the point sources detected in the field
and the Rotation Measure ($R\!M$).
Section~\ref{powTrSec} discusses the power transfer from
large to small angular scales
as a result of randomization of polarization angles. It
considers how this changes the polarized power spectrum.
The $E$- and $B$-mode power spectra of the synchrotron emission in
the patch will be presented in Section~\ref{APSSec} together with an analysis of
the role of Faraday rotation effects. The implications for
CMBP investigations are discussed in Section~\ref{cmbpSec}.
Summary and conclusions are finally presented in Section~\ref{concSec}.

\section{Observations}\label{obsSec}

The region centred on
($\alpha = 5^{\rm h}$, $\delta = -49^\circ$)
has been observed at 1.4~GHz by~\citet{be03}
using the Australia Telescope Compact Array
(ATCA, \citealt*{frater92}), an East-West synthesis
interferometer situated near Narrabri (NSW, Australia),
operated by CSIRO-ATNF.

The observation was performed as a 49 pointing mosaic covering
a $3^\circ \times 3^\circ$ region.
The array configuration used was the so-called EW214 including
spacing from 30~m to 214~m. This provides sensitivity
on angular scales ranging from $\sim 30$~arcmin down to the
resolution of $\sim 3.4$~arcmin.
The system provides the four Stokes parameters $I$, $Q$, $U$, $V$.
Details of the observations are listed in Table~\ref{obsTab} 
and presented in ~\citet{be03}.
\begin{table}
 \centering
  \caption{Main characteristics of the 1.4~GHz observations.}
  \begin{tabular}{@{}lr@{}}
  \hline
  Central Frequency &  1380~MHz \\
  Effective Bandwidth & $205$~MHz \\
  Array Configuration & EW214 \\
  Angular Sensitivity Range & 3.4--30~arcmin \\
  Location (J2000)& $\alpha = 5^{\rm h}$, $\delta = -49^{\circ}$ \\
  Area Size	& $3^{\circ}\times 3^{\circ}$ \\
  Observation Period & June 2002 \\
  Effective Observing Time & $\sim 70$~h \\
  Sensitivity (flux)& 0.18~mJy~beam$^{-1}$ \\
  Sensitivity (temperature)& 3.2~mK~beam$^{-1}$ \\
  \hline
  \end{tabular}
 \label{obsTab}
\end{table}

The maps of the Stokes parameters $I$, $Q$ and $U$, along with
the polarized intensity $I^p = \sqrt{Q^2+U^2}$,
are shown in Figure~\ref{iquFig}. Differing from~\citet{be03}, here no subtraction
of polarized point sources has been performed.

\begin{table*}
\begin{center}
\centering
\caption{Position, total intensity ($I$) and polarized intensity ($I^p$) of
	 the 9 sources detected in the field.
	 The rms-error is the same for both the two intensities
	 and corresponds to the beam-sensitivity (0.18~mJy~beam$^{-1}$).
	 Polarization angle ($\phi$) and polarization degree ($\Pi = I^p/I$) 
         are also reported. Counterparts were located in radio catalogues.
         Where no radio-counterpart was found, the closest optical object
         found in both the APM and 2-MASX surveys is listed
         (see text for details).
	 The last column provides the distance from the counterpart.}
\begin{tabular}{|c|c|c|c|c|c|c|l|c|}
source & RA & DEC & $I$ & $I^p$ & $\phi$ & $\Pi$ &
counterpart & distance\\
       & J2000 & J2000 & [mJy] & [mJy] &    &	 &    & [arcmin]\\
\hline
1	& $5^{\rm h} 01^{\rm m} 01^{\rm s}.4$ & $-48^\circ 31^\prime 03^{\prime\prime}.0$
& 164.10 & 10.72 & $15.4^\circ \pm 0.5^\circ$  &6.5\%
& PMN J0501-4831 & 0.5\\
2	& $5^{\rm h} 05^{\rm m} 58^{\rm s}.8$ & $-49^\circ 11^\prime 43^{\prime\prime}.0$
&  98.52 &  6.63 & $21.5^\circ \pm 0.8^\circ$  &6.7\%
& 2MASX J05055521-4910485 & 1.1\\
3	& $4^{\rm h} 58^{\rm m} 34^{\rm s}.5$ & $-48^\circ 34^\prime 40^{\prime\prime}.2$
& 130.20 &  3.71 & $43.8^\circ \pm 1.4^\circ$  &2.9\%
& APMUKS(BJ) B045704.24-483820.7 & 1.8\\
4	& $5^{\rm h} 00^{\rm m} 38^{\rm s}.2$ & $-49^\circ 12^\prime 29^{\prime\prime}.5$
& 139.70 &  3.40 & $43.6^\circ \pm 1.5^\circ $ &2.4\%
& PMN J0500-4912 & 0.2\\
5	& $4^{\rm h} 59^{\rm m} 37^{\rm s}.8$ & $-48^\circ 43^\prime 13^{\prime\prime}.6$
&  38.21 &  2.78 & $41.3^\circ \pm 1.9^\circ$  &7.3\%
& APMUKS(BJ) B045822.87-484738 & 0.9\\
6	& $5^{\rm h} 07^{\rm m} 08^{\rm s}.8$ & $-49^\circ 29^\prime 32^{\prime\prime}.0$
&  20.73 &  2.75 & $-27.5^\circ \pm 1.9^\circ$ &13.3\%
& 2MASX J05070829-4929268 & 0.1\\
7	& $4^{\rm h} 57^{\rm m} 17^{\rm s}.7$ & $-48^\circ 34^\prime 58^{\prime\prime}.5$
&  23.93 &  2.32 & $42^\circ \pm 3^\circ$      &9.7\%
& APMUKS(BJ) B045558.02-483812.5 & 1.3\\
8	& $5^{\rm h} 07^{\rm m} 24^{\rm s}.5$ & $-50^\circ 13^\prime 14^{\prime\prime}.4$
& 123.70 &  2.18 & $40^\circ \pm 2^\circ$      &1.8\%
& PKS 0506-502 & 0.2 \\
9	& $4^{\rm h} 59^{\rm m} 29^{\rm s}.7$ & $-48^\circ 31^\prime 56^{\prime\prime}.9$
&  55.48 &  2.01 & $44^\circ \pm 3^\circ$      &3.6\%
& PMN J0500-4912 & 0.3 \\
\hline
\label{source_positions}
\end{tabular}
\end{center}
\end{table*}

The polarized emission $I^p$ is {\it patchy} in nature, and distributed
over the whole field. An exception is a bright feature
in the South-West corner of the area: it is more
filamental, being about $1^\circ$ long.
The Stokes $I$ image does not show any particular diffuse structure
at the same coordinates, although the point source
contamination makes the comparison hard.
As in other high resolution observations (e.g. \citealt{wieringa93,gaensler01}),
a {\it Faraday screen} is likely acting along the line of sight.
This acts as a small scale modulation of a relatively uniform background,
generating the
apparent filamental structure on small angular scales.

The mean polarized emission
$P_{\rm rms} = \sqrt{\left<Q^2\right> + \left<U^2\right>} = 11.6$~mK
found by~\citet{be03} is well above the beam
sensitivity ($S/N \sim 3.5$) allowing an analysis of the synchrotron
emission features in the area.

\section{Point Sources}\label{pointSec}

Even though the field has been selected to minimize their contamination,
a few point sources are present in the polarized intensity map (Figure~\ref{iquFig}).
To identify them, we fit the maxima/minima in $Q$ and $U$ maps with a
2D-Gaussian beam and a constant value using MIRIAD's task IMFIT \citep*{sault95}. 
The constant value is adopted to account for the background emission.
We keep only those sources where the resultant Full Width at Half Maximum (FWHM)
is compatible with the synthesized beam.
Sources providing larger FWHM are instead discarded, 
the fit being not able to separate the background emission.

For the detections, we retain only those point sources where the emission
is 3~times the rms signal of the diffuse component ($S/P_{\rm rms} > 3$).
This gives good confidence that we avoid confusion
between point sources and statistical fluctuations of the diffuse background,
while ensuring a solid detection above the instrumental noise ($S/N > 10$).
Moreover, this criterium allows us to provide a complete list down to the
polarized flux of $S^p_{\rm lim} = 2.0$~mJy.

Given to these criteria, we find 9 sources: these
are listed in Table~\ref{source_positions}.

We have attempted to identify the sources in published catalogues
such as the Parkes-MIT-NRAO (PMN, \citealt{griffith93}) survey.
PMN was performed at 4.85~GHz, so that, assuming a mean spectral index
$\alpha = 0.7$, its $S_{4.85} = 42$~mJy flux-limit corresponds to
about $S_{1.4} \sim 100$~mJy
at 1.4~GHz. Accepting this, we would expect sources 1, 2, 3, 4 and 8
to be present in the PMN catalogue.
However only sources 1, 4 and 8 were found: we are not able to find any
identification for sources 2 and 3. It is worth noting that
their fluxes are near the catalogue limit:
a steeper spectrum would result in PMN failing to detect them.
Conversely, source~9 has a PMN
counterpart, even if the measured 1.4~GHz flux is about a half of the
limit extrapolated from 4.85~GHz.

When a PMN counterpart could not be found, we looked for the closest object
present in the optical/IR catalogues APM \citep{maddox90} and
2-MASX \citep{huchra03},
provided it lies within a FWHM of the detected radio source.
Possible counterparts have been found for all the missing
objects.
These are listed in Table~\ref{source_positions}.
It is worth noting that all of these sources are classified as normal galaxies.

Given the limited number of sources,
no general properties can be extracted. We simply note
that only one source exceeds the 10\% polarization.

\section{Rotation Measure}\label{rmSec}

The magnetic field parallel to the line of sight
($B_\parallel$) changes the polarization angle by Faraday rotation.
The variation $\Delta \phi = \phi - \phi_0$ with respect to the intrinsic
polarization angle $\phi_0$ is given by the formula
\begin{equation}
      \Delta\phi = R\!M \; \lambda^2,
\end{equation}
where $\lambda$ is the wavelength of the radiation and $R\!M$
is the Rotation Measure. The latter is defined by the integral
along the line of sight
\begin{equation}
      R\!M = 0.81\int B_\parallel(l)\;n_e(l)\;dl
\end{equation}
which depends on  
$B_\parallel(l)$ and the free-electron density $n_e(l)$
in the Interstellar Medium (ISM) at various distances $l$ from the observer.

Beside changing in the polarization vector direction and, in turn,
in the naive estimate of the magnetic field orientation, Faraday
rotation can induce both depolarization effects and
Faraday screen modulations. These
can significantly modify the angular power distribution and pattern of the
polarized emission.
$R\!M$ estimates in the patch of sky being consider are thus important 
in understanding the
significance of these effects.

Because the ATCA correlator produces a number of frequency channels
across the observed bandwidth, we are able to determine
values of $R\!M$ from our data set.
Faraday rotation has been evaluated by grouping the 26
useful channels of our observations
in four sub-bands to
form four maps of $Q$, $U$ and $V$ at the
different frequencies of 1316, 1368, 1404 and 1454~MHz.
For each frequency, the Stokes $V$ map sets the noise level for the
corresponding $Q$ and $U$ maps.
Using MIRIAD's task IMPOL,
we have obtained maps of polarized
intensity and polarization angle at each frequency.
$R\!M$ values are then determined using MIRIAD's task IMRM, that
fits a $R\!M$ map from polarization angles at the different
frequencies. This task
also tries to solve the $N\pi$ ambiguity in the $R\!M$ determination.
Pixels with an error in polarization
angle greater than $20^\circ$ are discarded.

\begin{figure}
  \includegraphics[angle=0, width=1.0\hsize]{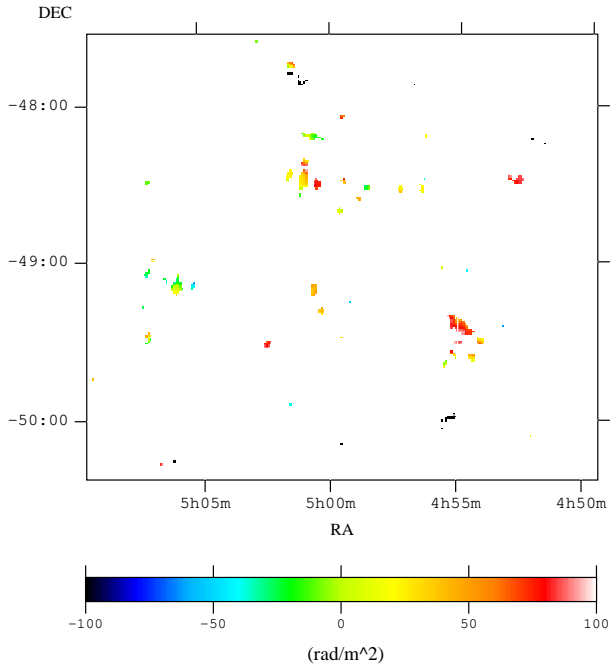}
  \caption{Map of $R\!M$s measured in the patch.}
  \label{rmMapFig}
\end{figure}

\begin{figure}
  \includegraphics[angle=0, width=1.0\hsize]{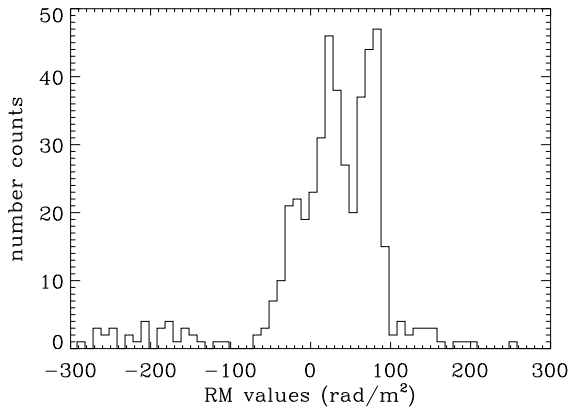}
  \caption{Distributions of the measured $R\!M$s.}
  \label{rmHistFig}
\end{figure}

The $R\!M$ values that were successful determined by
this process are shown in
Figure~\ref{rmMapFig}.
The coverage of the measurements in the patch is sparse, and so
they cannot be considered
to be fully representative of the area.
However, they are scattered over all the field
and thus give a useful indication of the broader trend.
The distribution,
shown in Figure~\ref{rmHistFig}, presents two peaks: the first
near 20~rad~m$^{-2}$ and a second 
near 80~rad~m$^{-2}$.
Apart from a small number of points,
the measurements of the second peak relate to the filament structure.
This is thus characterized by $R\!M$ values larger than typical of this 
patch. This might be expected if the filament is
caused by a Faraday screen.
The other measurements of $R\!M$ lie near the first peak.
Thus 20~rad~m$^{-2}$ can be considered a typical value of the
$R\!M$ of the successful pixels.

This result is in good agreement with other
estimates of $R\!M$ in this region.
The all-sky catalogue of rotation measures for extragalactic sources by
\citet*{broten88}
contains four sources within $5^\circ$ from the centre of the area.
These have $R\!M$ values of 13, 28, 34 and 53~rad~m$^{-2}$.
The rotation measure maps of \citet{han97} and \citet*{john04} are
indicative of large-scale $R\!M$ values.
In our area, they suggest values of approximately 20--30~rad~m$^{-2}$
(see~\citealt{han04} for a review).

Given these various estimates of $R\!M$, we can consider
$R\!M = 50$~rad~m$^{-2}$
as a reasonable upper limit for the typical rotation measure within the
patch.
As in \citet{be03}, this value in combination with
the total bandwidth of the ATCA observations implies
a depolarization factor of
$D\sim 92$\%. This has a marginal impact on our main goal of
estimating the level of polarized synchrotron emission in the area.

\section{Power transfer from large to small scales}\label{powTrSec}

As noted by several authors (e.g. see~\citealt{wieringa93} and
~\citealt{gaensler01}), Faraday screens can
introduce a small scale modulation of a relatively uniform background.
If a complex Faraday screen is interposed between
the observer and a uniform polarized emission,
the $R\!M$ pattern of the screen
will cause    
modulation of polarization angle, resulting in
variable patterns of $Q$ and $U$.
This causes transferring power from large to small
angular scales in polarization maps, and generates 
{\it false} small scale structures.

\begin{figure}
  \includegraphics[angle=0, width=1.0\hsize]{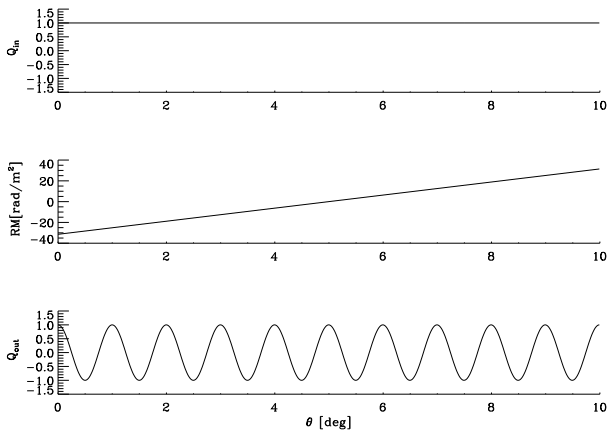}
  \caption{A simple model showing how a variable $R\!M$ pattern can transform
	   uniform emission to one modulated on smaller angular scales.
	   The top panel shows a background with uniform emission
	   on a $10^\circ$ box-size
	   (emission of Stokes $Q$ only is assumed, i.e. $\phi_0 = 0^\circ$).
	   The mid panel shows $R\!M$ varying linearly.
	   The {\it observed} $Q$ is shown in the bottom panel,
	   assuming a 300~MHz frequency. The {\it observed} emission
	   is modulated on $1^\circ$ scale, smaller than the intrinsic
           emission.}
  \label{toyRMFig}
\end{figure}

Figure~\ref{toyRMFig} demonstrates this point using a simple model:
the intrinsic emission is uniform, while the observed one is sinusoidal.
This is caused by a $R\!M$ which varies linearly with angle.
A result of the process is that angular power that had corresponded to
the large scale structure has been transferred down to
the modulation scale.

Overall the effect randomizes the polarized emission larger than
a particular angular scale.

Although this effect has been invoked as a qualitative
explanation of structures observed, we are not aware
of a quantitative analysis having been performed.

Here we simulate some realistic examples to estimate
importance and main features of the effect.
First, we generate $Q$ and $U$ maps
starting from $E$- and $B$-mode angular power spectra which
we assume have power law behavior
\begin{eqnarray}
      C^E   &\propto&	\ell^{\beta_E}, \nonumber\\
      C^B   &\propto&	\ell^{\beta_B},
      \label{cecbSimEq}
\end{eqnarray}
with the spectral indeces which can be
\begin{equation}
   \beta_{X} = 0.0,\, -1.5,\, -3.0, \hskip 1cm X=E,B.
\end{equation}
These have been selected in a range wide enough to cover all of the cases observed
to date for the polarized Galactic synchrotron emission and are centred
on the most common measured values ($\beta_{X} = -1.6$, e.g. see
 \citealt{br02,gi02}).
\begin{figure}
  \includegraphics[angle=0, width=1.0\hsize]{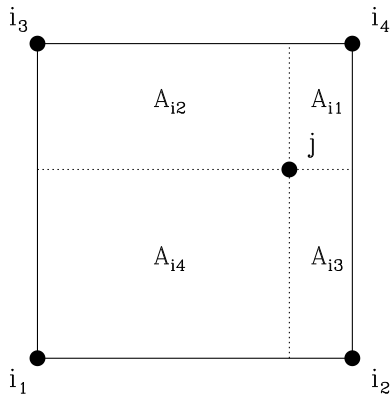}
  \caption{CIC scheme for regularly gridded data. The four grid points
	   $i_1$, $i_2$, $i_3$ and $i_4$ are the nearest to the pixel $j$ where
	   the quantities sampled on the grid points have to be interpolated.
	   $A_{i_k}$ with $k=1,4$ are the opposite sub-cells used in the
	   weighted mean.}
  \label{CICFig}
\end{figure}

We use the \verb"synfast" procedure of the HEALPix package \citep*{go99}
to generate the maps given the power spectra.
We adopt an angular resolution of about 2~arcmin
(HEALPix's parameter $N_{\rm side} = 2048$) to match the angular resolution
of the observed maps (about 3.4~arcmin).
Finally, $20^\circ\times20^\circ$ square maps are extracted.

The area we observed shows
a polarization angle pattern uniform at least up to 10--15~arcmin
scale (see~\citealt{be03}). 
In this simulation, we
assume that the polarization angles are totally
random above 15 arcmin angular scale size.
This is done by placing a 15~arcmin~$\times$~15~arcmin
grid on the simulated map, and then
assigning a random polarization rotation angle to each grid point.

To associate a random angle with each pixel of the map,
the random angles of the nearest grid points are linearly interpolated
to the pixel position by using a Cloud-in-Cell (CIC) scheme~\citep{he81}.
In order to explain the procedure, let us define
$\theta_i$ as the random angle associated to the
$i^{\rm th}$ point of the grid and $j$ the $j^{\rm th}$ pixel of the map
where the angles have to be interpolated.
Referring to Figure~\ref{CICFig}, the CIC method consists first
in finding the 4 grid points $i_1$--$i_4$ which are
the nearest to the pixel $j$ and which define the Cell where $j$ itself is located.
The linear interpolation is performed assigning to $j$ a weighted mean
of the values sampled on these 4 points:
the pixel $j$ divides
the Cell in four sub-cells and the weight of a grid point is
proportional to the area of the sub-cell opposite to the point itself:
\begin{equation}
 \theta_j = {\sum_{k=1}^4 A_{i_k}\,\theta_{i_k} \over \sum_{k=1}^4 A_{i_k}},
 \label{cic1Eq}
\end{equation}
where $\theta_j$ is the quantity to be estimated at the pixel $j$,
$i_k$ with $k=1,4$ are the four nearest grid points and $A_{i_k}$ are the
areas of the corresponding sub-cells.

We have adopted this linear interpolation scheme to avoid the discontinuities
that a simple Nearest Grid Point (NGP) scheme would produce in the
polarization angle map.
\begin{figure}
  \includegraphics[angle=0, width=1.0\hsize]{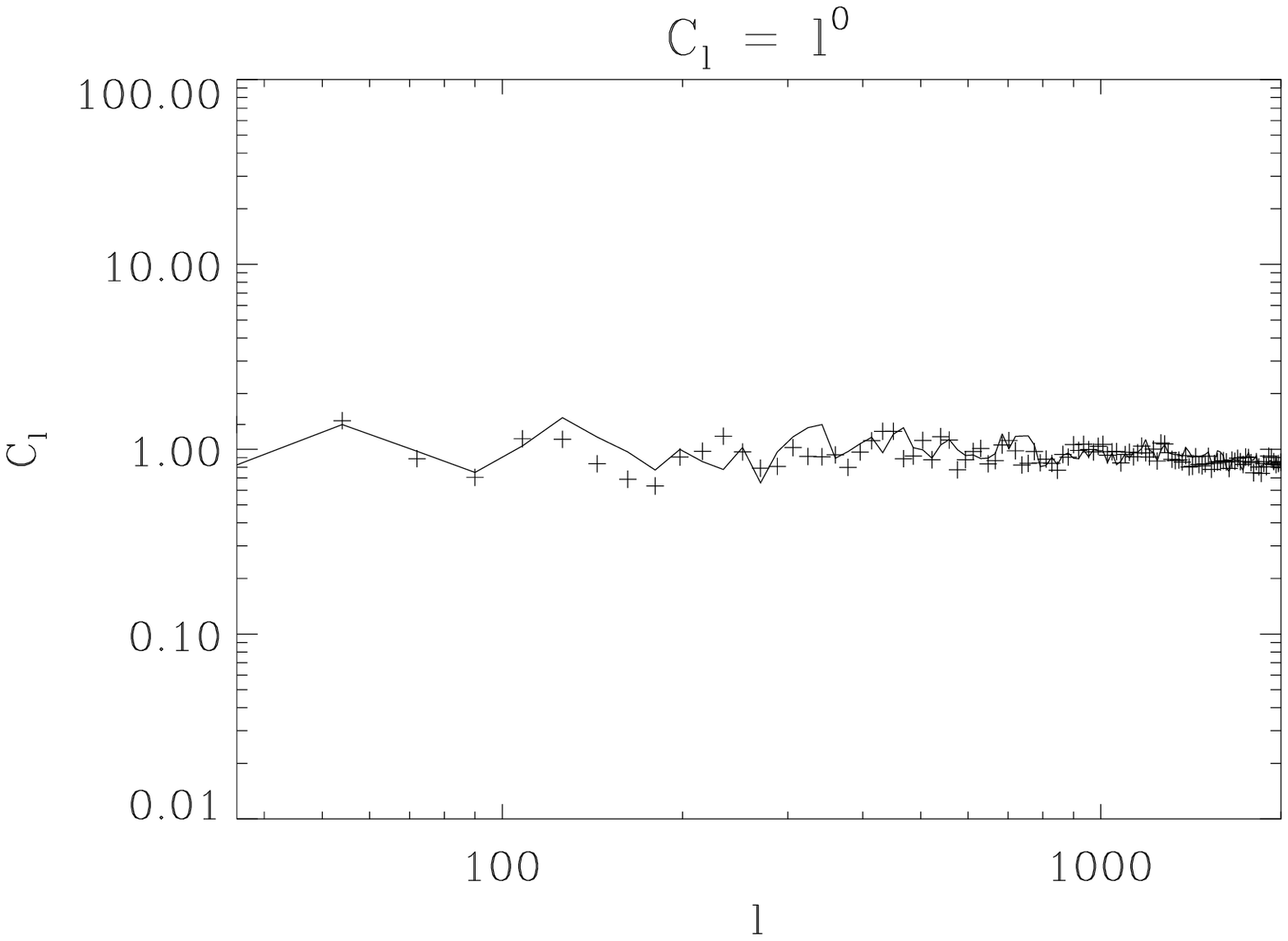}
  \includegraphics[angle=0, width=1.0\hsize]{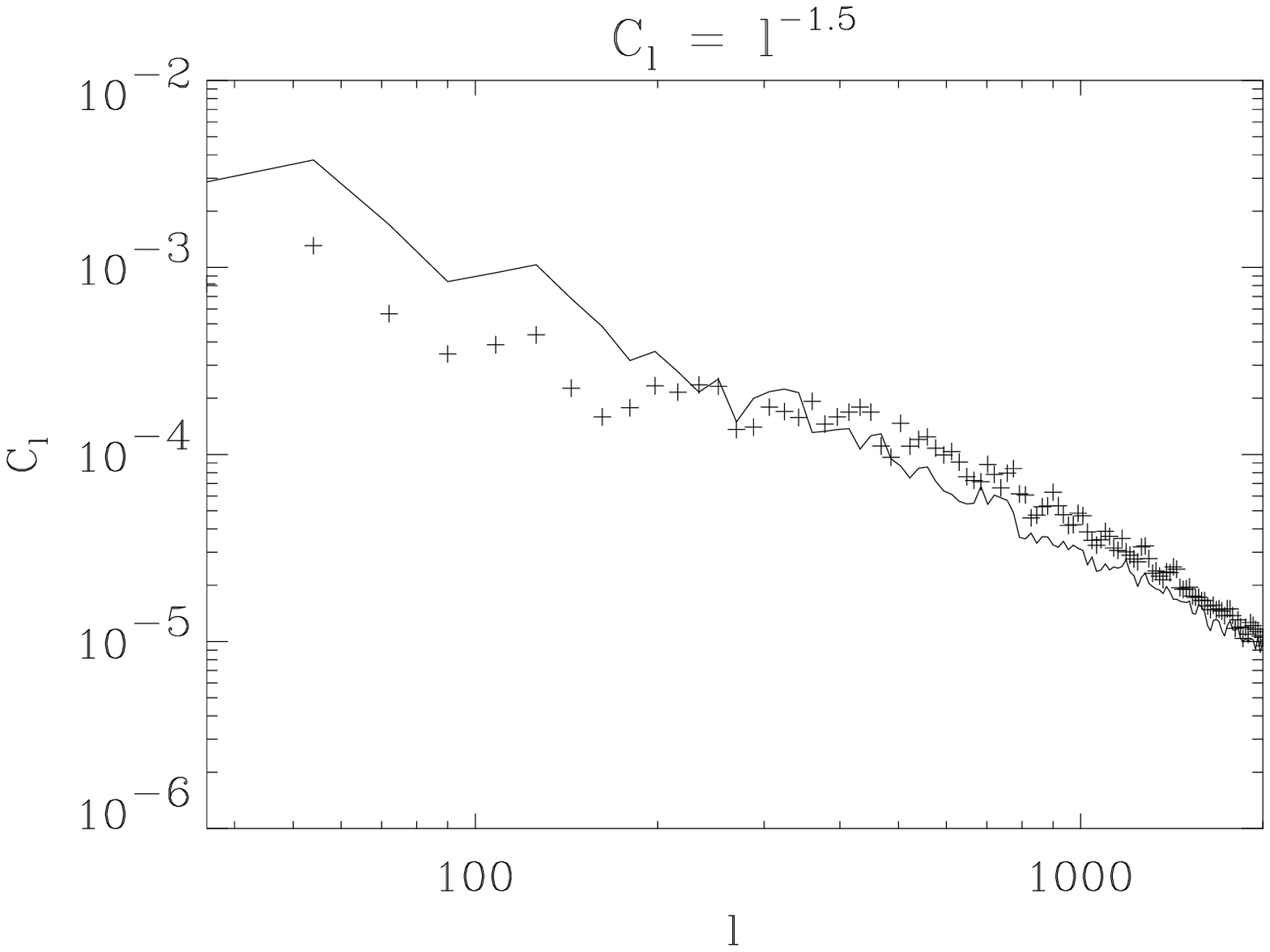}
  \includegraphics[angle=0, width=1.0\hsize]{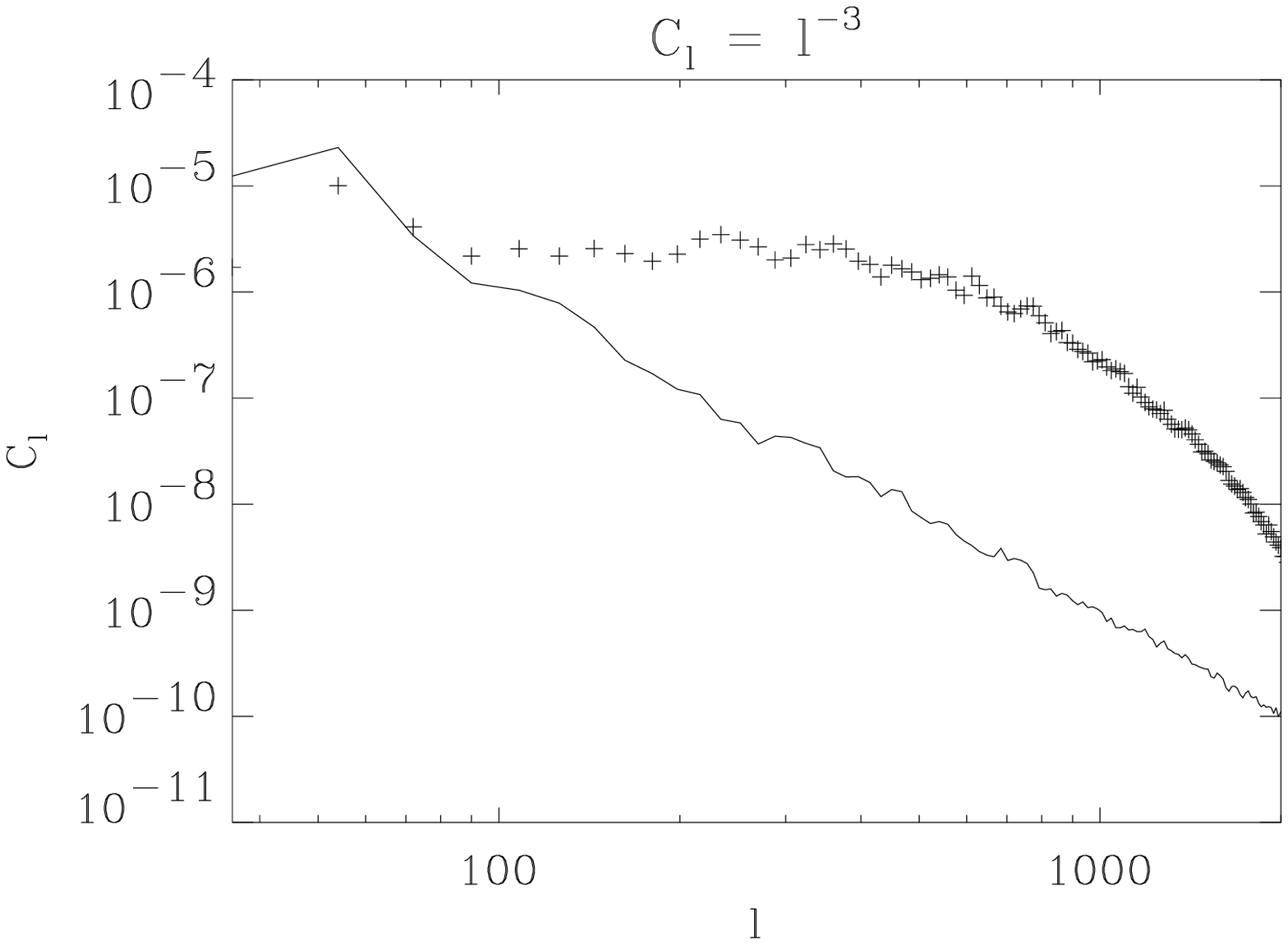}
  \caption{$C^E$ power spectra for the three models described in the text
  ($\beta_E =$~0.0,~~-1.5,~~-3.0) before (solid lines) and after (crosses)
  the application of the polarization angle randomization procedure.}
  \label{randPSFig}
\end{figure}

Figure~\ref{randPSFig} shows the power spectra $C^E$ before and after the
randomization procedure. The results for $C^B$ are similar.
For the case $\beta_E = 0.0$, there is no change in the power at any scale.
This is expected as this case corresponds to 
a totally random pattern of $Q$ and $U$ ($C_\ell = $~{\it constant} is the case 
of pure white noise):
adding further randomization does not change the statistics of the data.

In the other two cases the randomization reduces the power
on the largest scales, while increasing it on the smallest ones.
In particular, on the scales that the interferometer is sensitive to
(3--30~arcmin), the power is greater or equal to that of the input maps.

Randomization of the large-scale polarization angle thus does
transfer power
from large to small angular scales. On the small scales, the angular
power detected is an
enhancement. Consequently what we measure is an upper limit to the intrinsic
fluctuation.

The angular scale which marks the transition from reduction to enhancement
depends on the power law index. However, at least for
the range explored here, it is larger than the scale size of the randomization.
Because our map has a randomization scale of  at least 15~arcmin, the power
detected in the 3--30~arcmin range is enhanced or at least not reduced.
Consequently, in terms of power spectra, our measurements represents
an upper limit of the real emission.

From Figure~\ref{randPSFig}, we can consider the slope of the
power spectrum:
this power spectrum is steeper in those 
$\ell$-ranges where the power is enhanced.
Indeed, apart from a transition range centred at
the randomization scale, the modified spectra follows a power law
but with a steeper slope. This is in agreement with the results of
\citet{tu02}. They found spectra at 1.4~GHz that were
steeper than at 2.4~GHz, where Faraday effects are weaker.

\section{Power Spectrum Analysis}\label{APSSec}

$E$- and $B$-modes,
combinations of the tensorial 2-spin quantities
$Q\pm j\,U$, completely describe the polarized emission and
have the useful characteristic of being scalar
(e.g. see \citealt{zaldarriaga97,zaldarriaga98}).
Thus, they allow us to
describe how the polarized signal is distributed
across angular scales by using simple
scalar spherical harmonics.

We prefer these descriptors instead of the scalar spectra
of $Q$ and $U$ because the latter depend on the orientation
of the reference frame.
Additionally, $E$- and $B$-spectra are the quantities predicted
by cosmological models.
Their use for Galactic work allows a direct comparison between
the Galactic and cosmological signals and
the evaluation of the contamination
of the latter by the former.

\begin{figure}
  \includegraphics[angle=0, width=1.0\hsize]{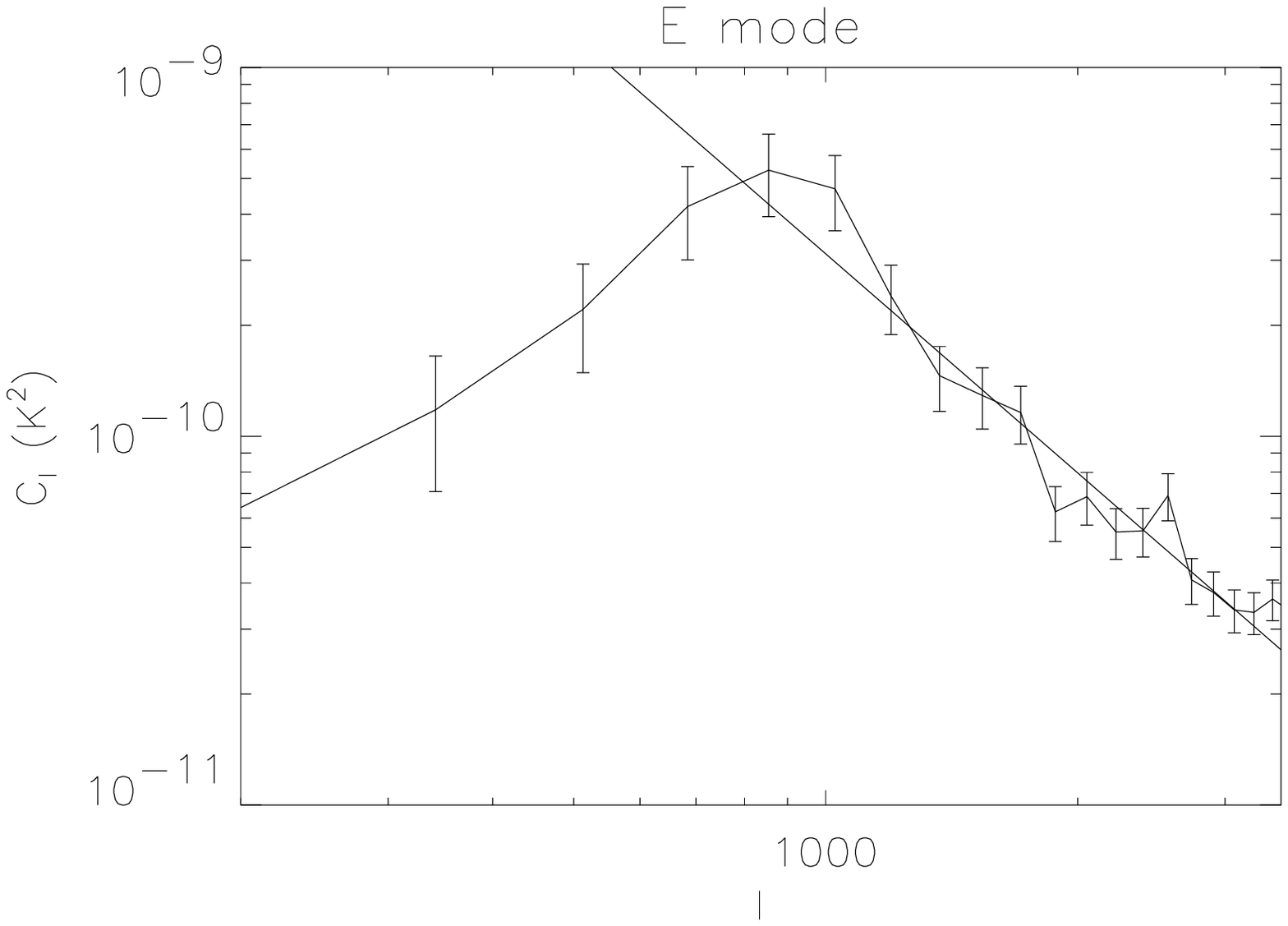}
  \includegraphics[angle=0, width=1.0\hsize]{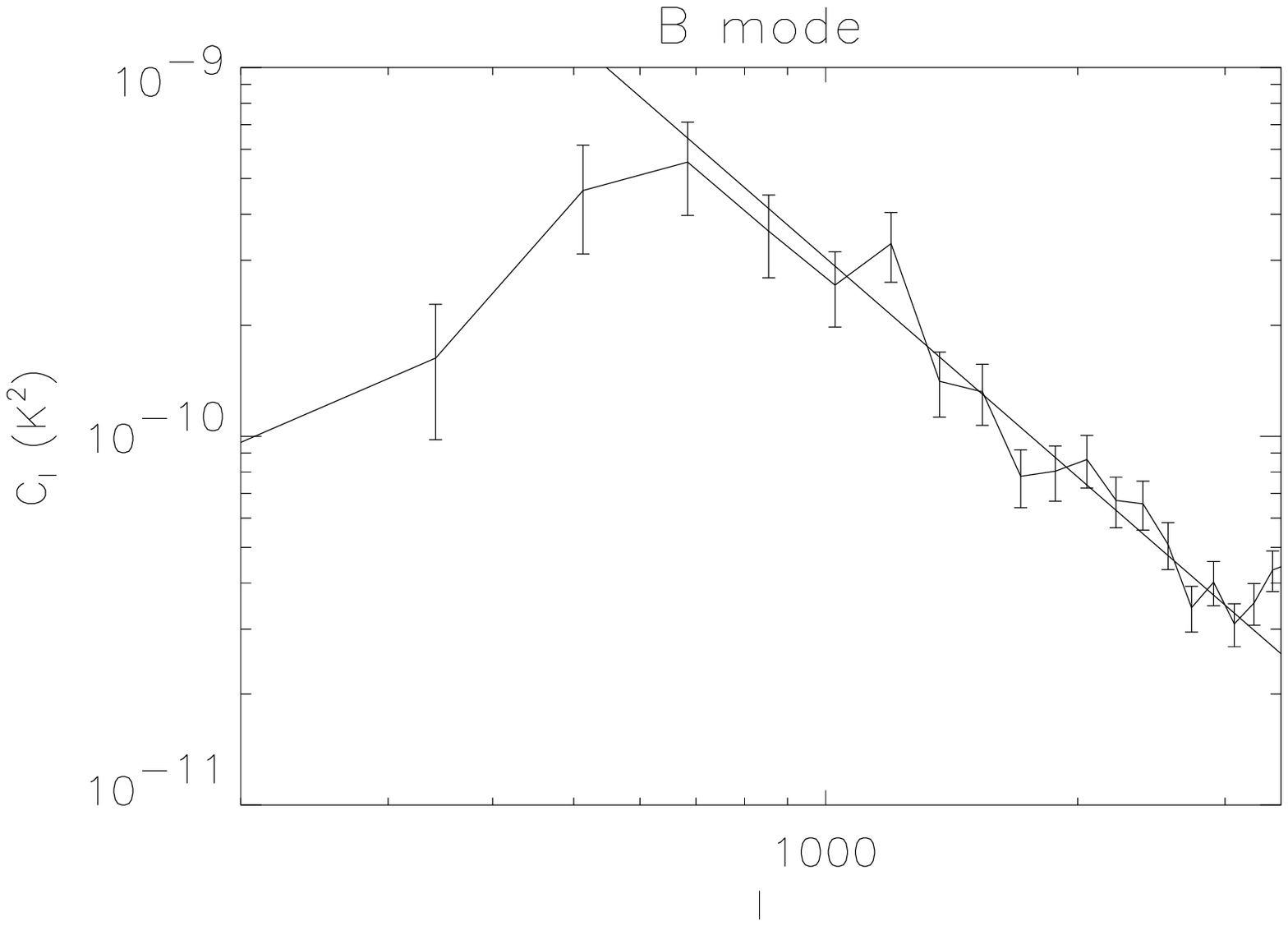}
  \caption{$E$- (top) and $B$-mode (bottom) angular power spectra of
  the polarized emission at 1.4~GHz in the observed patch.}
  \label{boomSpecFig}
\end{figure}

We have computed the $E$- and $B$-mode spectra by using the Fourier technique
of~\citet{seljak97}. The results are shown in Figure~\ref{boomSpecFig}.
Both power spectra are well approximated by a power law
on scales smaller than about 15~arcmin (multipoles $\ell > 800$).
On larger scales, there is a  turn-over and power
becomes negligible for  $\ell < 300$--$400$. This
corresponds to about the 30~arcmin antenna primary beam size.
This turn-over is a result of the spatial filtering of an interferometer.
The ATCA has
little sensitivity on scales larger than about the FWHM of
the primary beam, has somewhat improved sensitivity between
about FWHM and FWHM$/2$,
and shows full sensitivity on scales smaller than about FWHM$/2$.

The power law behavior covers the 3.4--15~arcmin range
of full sensitivity of the instrument. Table~\ref{powFitTab} gives the
results when we fit the functional form
\begin{equation}
   C^{X}_{\ell} = C_{2000}^{X}\,\left({\ell\over 2000}\right)^{\beta_X},
		  \hskip 1.0cm X=E,B,
\end{equation}
in the $\ell$--range of 800--2800.
\begin{table}
 \centering
  \caption{Best fit parameters of the
	   angular power spectra of our 1.4~GHz data.}
  \begin{tabular}{@{}lcc@{}}
  \hline
   Spectrum	&   $C^X_{2000}$\,[$10^{-11}$~K$^2$]	 &   $\beta_X$	      \\
  \hline
   $C^E_\ell$	     &	 $8.0\,\pm\,0.2$   &  $-1.97\,\pm\,0.08$ \\
   $C^B_\ell$	     &	 $8.0\,\pm\,0.2$   &  $-1.98\,\pm\,0.07$ \\
  \hline
  \end{tabular}
 \label{powFitTab}
\end{table}
\begin{table}
 \centering
  \caption{Slopes $\beta_X$ of angular power spectra of the Galactic
	   synchrotron polarized emission.
	   The table also reports the sky area the slope has been computed for and the
	   observation frequency.
	   SGPS--2.4 refers to the slope computed from the 2.4~GHz
	   survey \citep{duncan97} in the area covered by the SGPS
	   Test Region.
	   The last column reports the references where the slopes
	   have been measured.}
  \begin{tabular}{@{}lccc@{}}
  \hline
   Area     &	$\beta_X$   &	frequency & Ref.    \\
  \hline
   This area  &  $\sim -2.0$  &  1.4~GHz      & this work \\
   SGPS 			     &	$\sim -2.8$   &  1.4~GHz      &  \citet{tu02} \\
   SGPS-2.4			     &	$\sim -1.7$   &  2.4~GHz      &  \citet{tu02} \\
   Galactic plane      &  $\sim -1.6$	&  2.4--2.7~GHz &  \citet{br02} \\
  \hline
  \end{tabular}
 \label{slopeTab}
\end{table}

It is interesting to compare the amplitude and spectral index
with information from other Galactic surveys.
First, let us consider the SGPS Test Region \citep{gaensler01}.
This covers a similar area (about $4^\circ \times 6^\circ$).
Like the current study, the SGPS Test Region survey is an interferometric
observation at arcminute resolution and carried out at the same 1.4~GHz
frequency. However, its region of interest is located on the Galactic plane.

The first significant difference is in the amplitude:
SGPS has a $C_{2000}^X \sim 5\times 10^{-8}$~K$^2$~\citep{tu02},
corresponding to a signal about 25
times stronger than in the patch of the current study.
This is expected,
as the observed patch is at high Galactic latitudes. Furthermore
this is consistent
with the result of~\citet{be03}, which found that the mean
signal of this patch is about 10 times smaller than the background
emission in areas close to
the Galactic plane of the 1.4~GHz EMLS survey
\citep{uyaniker99}.
However it is worth noting that the difference
is not as large as the case with total intensity: in
total intensity
the differences between Galactic plane and high Galactic latitudes emissions
can be much larger -- for instance, see the full-sky surveys at
408~MHz~\citep{haslam82} and 1.4~GHz \citep*{reich86,reich01}.

Another area of significant difference is the slopes of the power spectra
(Table~\ref{slopeTab}).
The power laws in the observed patch have $\beta_X \sim -2.0$,
which is considerably flatter than
$\beta^{\rm SGPS}_X \sim -2.8$ for the SGPS~\citep{tu02}. However it is
close to the $\beta^{\rm SGPS-2.4}_X \sim -1.7$
found at 2.4~GHz in the region covered by the SGPS, but using the data of
\citet{duncan97} \citep{tu02}. The latter
agrees with the mean value $\bar{\beta}^{\rm 2.4-2.7}_X  \sim -1.6\,\pm\,0.2$
measured at 2.4--2.7~GHz over a large portion of the Galactic Plane
\citep{br02}.

The large difference between
$\beta^{\rm SGPS-2.4}_X$, $\bar{\beta}^{\rm 2.4-2.7}_X$ and
$\beta^{\rm SGPS}_X$ was first addressed
by~\citet{tu02}, who explained the steepening
of the SGPS through a Faraday-rotation induced transfer of
power from large to small scales.
Section~\ref{powTrSec} provides a quantitative approach
to this effect: Faraday screens can introduce randomization in the polarization
angle, which actually transfer power from large to small angular scales
and, thus, increases the power and the slope on scales smaller than
the randomization scale size.

Faraday rotation has a square dependence on the wavelength and
the effects at 2.4~GHz are expected to be less significant than at 1.4~GHz.
The flatter slope at 2.4~GHz can be interpreted as a reduction of
the effects of Faraday rotation. The slopes most probably
are near the intrinsic value.

Moreover, in spite of a large variation
of $R\!M$s across the Galactic plane, the slopes measured
at 2.4--2.7~GHz  have a small dispersion
($\Delta\beta^{\rm 2.4-2.7}_X = 0.2$),
indicating a good stability around the mean value.
This further supports the idea that the 2.4--2.7~GHz slopes are
slightly modified by Faraday effects and are
near the intrinsic values.

Our results are lie 
between the two previous values.
In fact, pulsar and extragalactic source measurements show that
the $R\!M$ becomes less significant at high
Galactic latitudes:
starting from the typical value $|R\!M| \sim 200$~$\rm rad\;m^{-2}$
at the Galactic Plane (e.g. ~\citealt{gaensler01}),
through 20--30~$\rm rad\;m^{-2}$ at mid Galactic latitudes \citep{han97,john04}
and down to 10--20~$\rm rad\;m^{-2}$ in the North Galactic Pole (NGP)
(e.g. see \citealt{sun04}).
Hence we would expect to see less impact of Faraday rotation 
in the observed patch than in the SGPS. 
We conclude that our observed patch is 
in an intermediate domain between
significant and negligible Faraday rotation effects.

We have analysed the data of~\citet{bs76}, which has led us to
a similar conclusion. In more detail,
the data of \citet{bs76} give polarization observations of about
a half of the sky at five frequencies: 408, 465, 610, 820, 1411~MHz.
Having the desirable characteristics of covering
all the Northern Galactic latitudes and a wide range in frequency,
these data allow us to explore behavior in both
frequency and latitude.
\begin{figure}
  \includegraphics[angle=0, width=1.0\hsize]{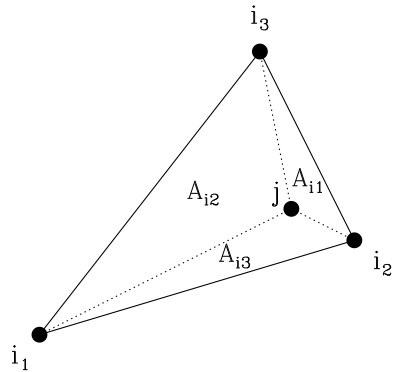}
  \caption{As for Figure~\ref{CICFig} but using the modified CIC scheme dealing
	   with irregularly sampled data (see text for details).}
  \label{CICTrFig}
\end{figure}
An issue with this data-set is that it is
sparsely sampled and arranged on irregular grids.
While the lowest frequency case is tolerable (the best
sampled areas have about $2^\circ$ sampling versus a FWHM~=~$2.3^\circ$)
the situation is worst at 1411~MHz,
where FWHM~=~$0.6^\circ$.
Nevertheless, regridding and smoothing on angular scales
larger than the sampling distance allow the analysis of
the characteristic of the emission at least for the large scale structure.

To account for irregular sampling, we regrid the data on regular
HEALPix maps via linear interpolation, using
a variant of the CIC
scheme. The standard CIC deals with data sampled on
a regular grid, as described in Section~\ref{powTrSec}.
On the other hand, here the data are sampled on irregular grids.
Let ($Q_i$, $U_i$) be the $i^{\rm th}$ sample of the grid
and $j$ the $j^{\rm th}$ pixel of the map to be interpolated.
Referring to Figure~\ref{CICTrFig}, we generalize the CIC scheme finding
the 3 grid points ($i_1$, $i_2$, $i_3$)
which define the smallest triangle containing $j$.
Connecting $j$ to $i_1$, $i_2$ and $i_3$, three triangular sub-cells
can be identified.
Similarly to Eq.~(\ref{cic1Eq}), the linear interpolation
is performed assigning to $j$ an average
of the values sampled on these 3 grid points weighted for the area
of the opposite sub-cell:
\begin{equation}
 X_j = {\sum_{k=1}^3 A_{i_k}\,X_{i_k} \over \sum_{k=1}^3 A_{i_k}},
 \hskip 1cm X=Q,U,
 \label{cic2Eq}
\end{equation}
where $X_j$ is the quantity to be estimated at the pixel $j$.

The deformations of the 2-sphere space can
generate depolarization due to the projection of $Q$ and $U$ onto the local
reference frame of parallels and meridians if a simple mean of the data
is performed. Close to the NGP, this will be a particular issue.
Following~\citet{br02}, to avoid these projection effects,
we perform a parallel transport of the polarization vectors
($Q_{i_k}$, $U_{i_k}$) onto the interpolation
point $j$ before averaging the data.

To account for the irregular sampling	distance,
the interpolated data are smoothed with a Gaussian filter of $4^\circ$~FWHM,
leading to maps able to describe the large scale distribution of the polarized
emission.

\begin{figure*}
  \includegraphics[angle=90, width=0.4\hsize]{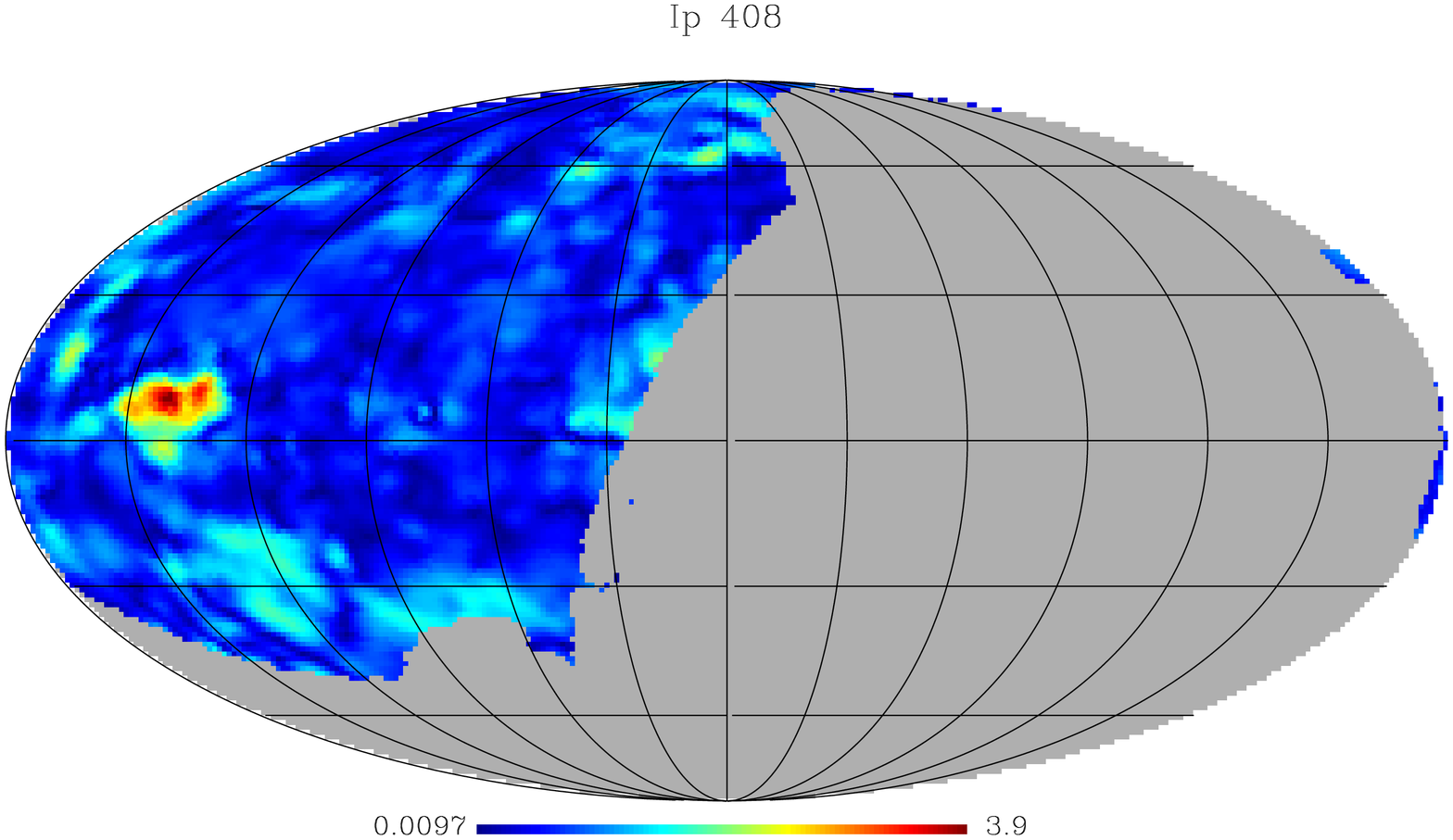}\hskip 1.0cm
  \includegraphics[angle=90, width=0.4\hsize]{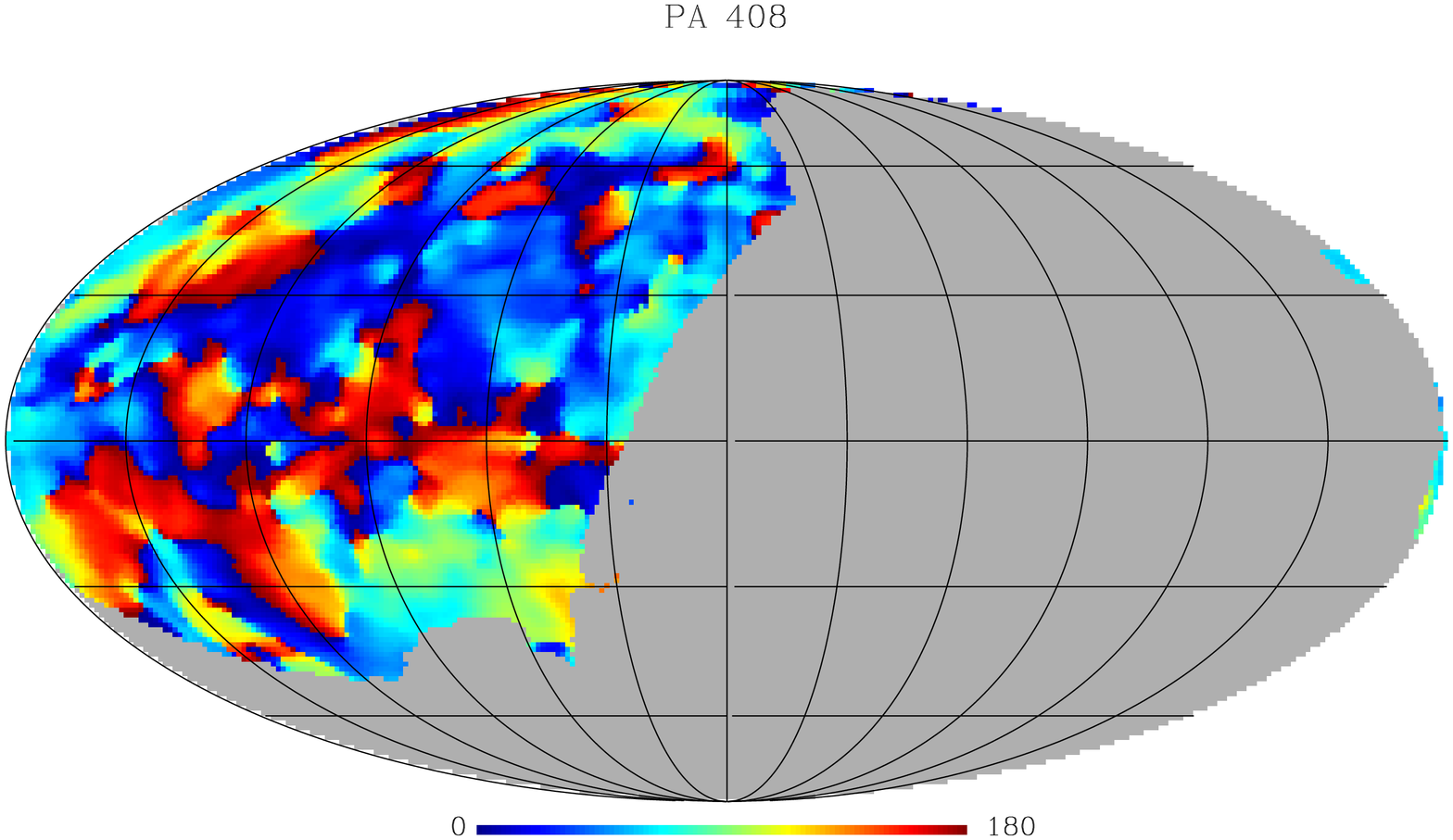}
  \includegraphics[angle=90, width=0.4\hsize]{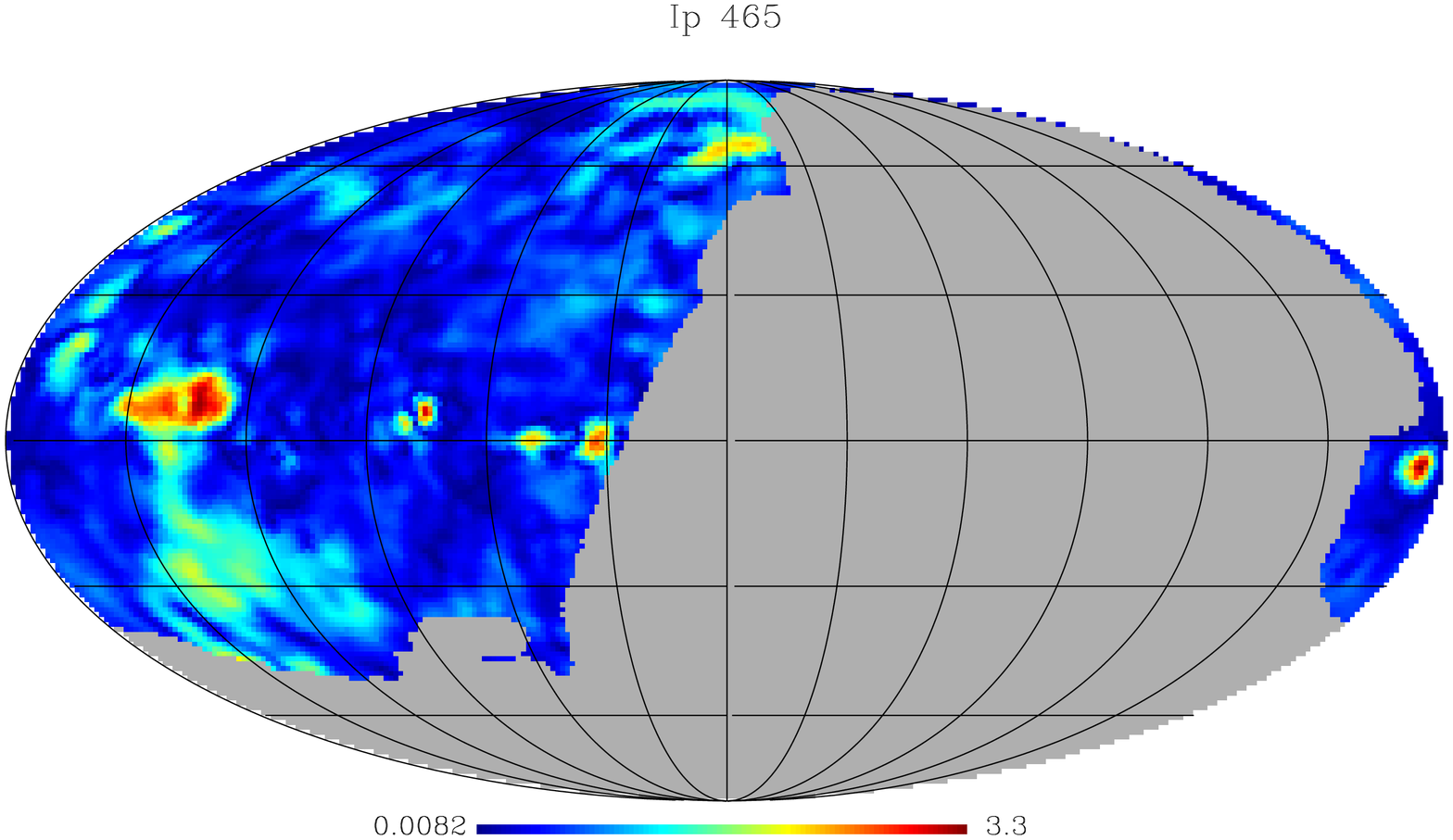}\hskip 1.0cm
  \includegraphics[angle=90, width=0.4\hsize]{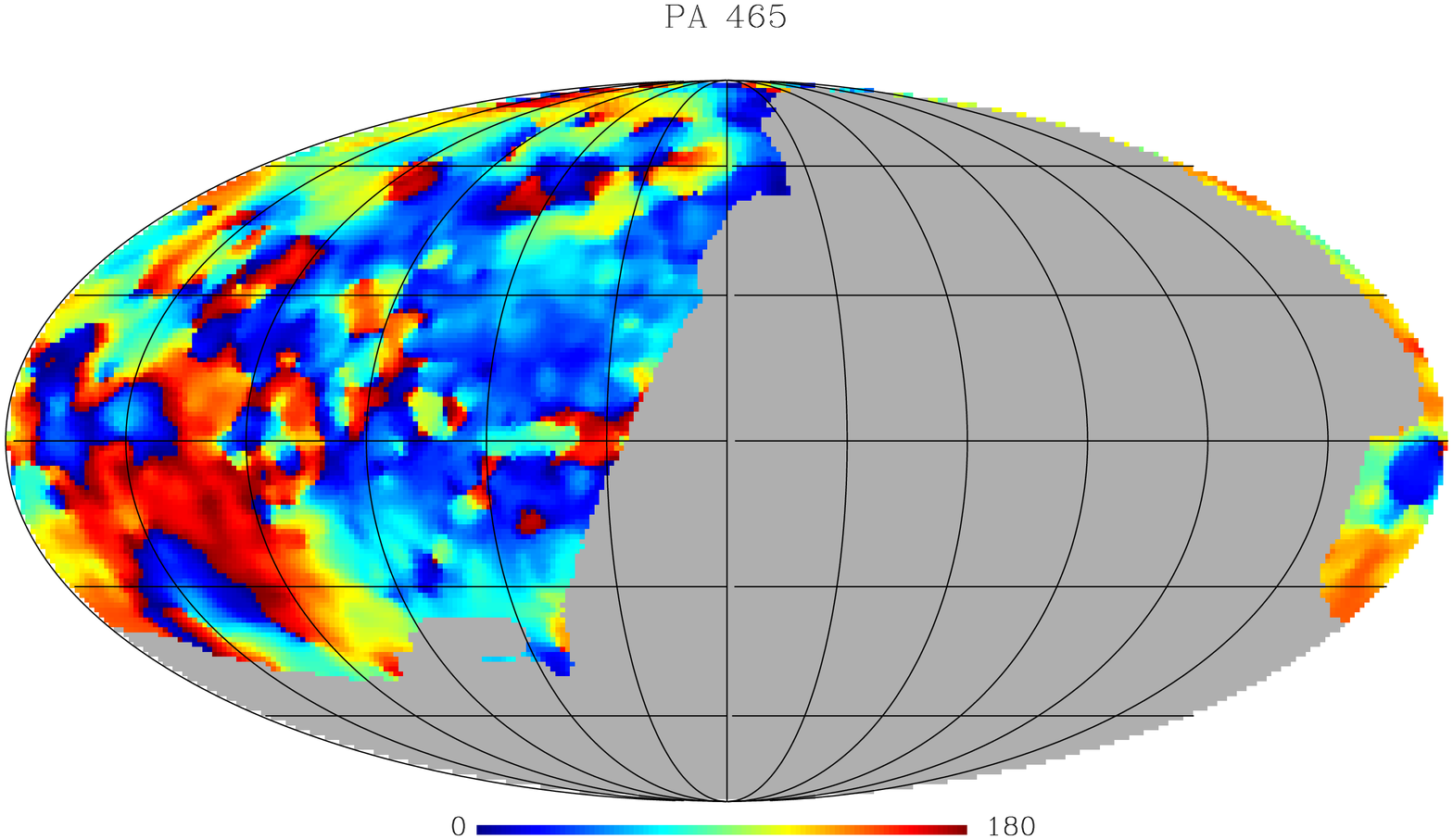}
  \includegraphics[angle=90, width=0.4\hsize]{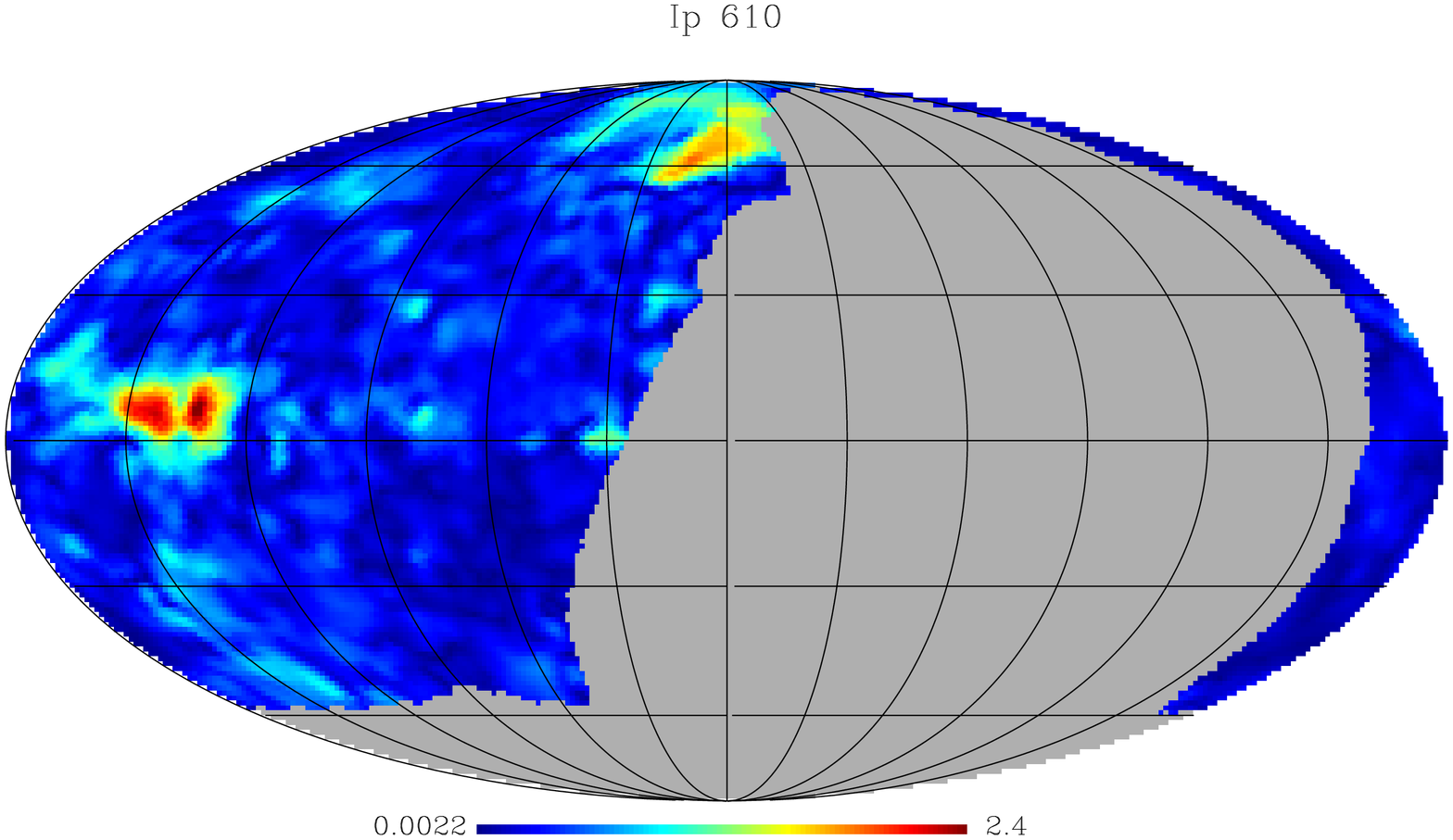}\hskip 1.0cm
  \includegraphics[angle=90, width=0.4\hsize]{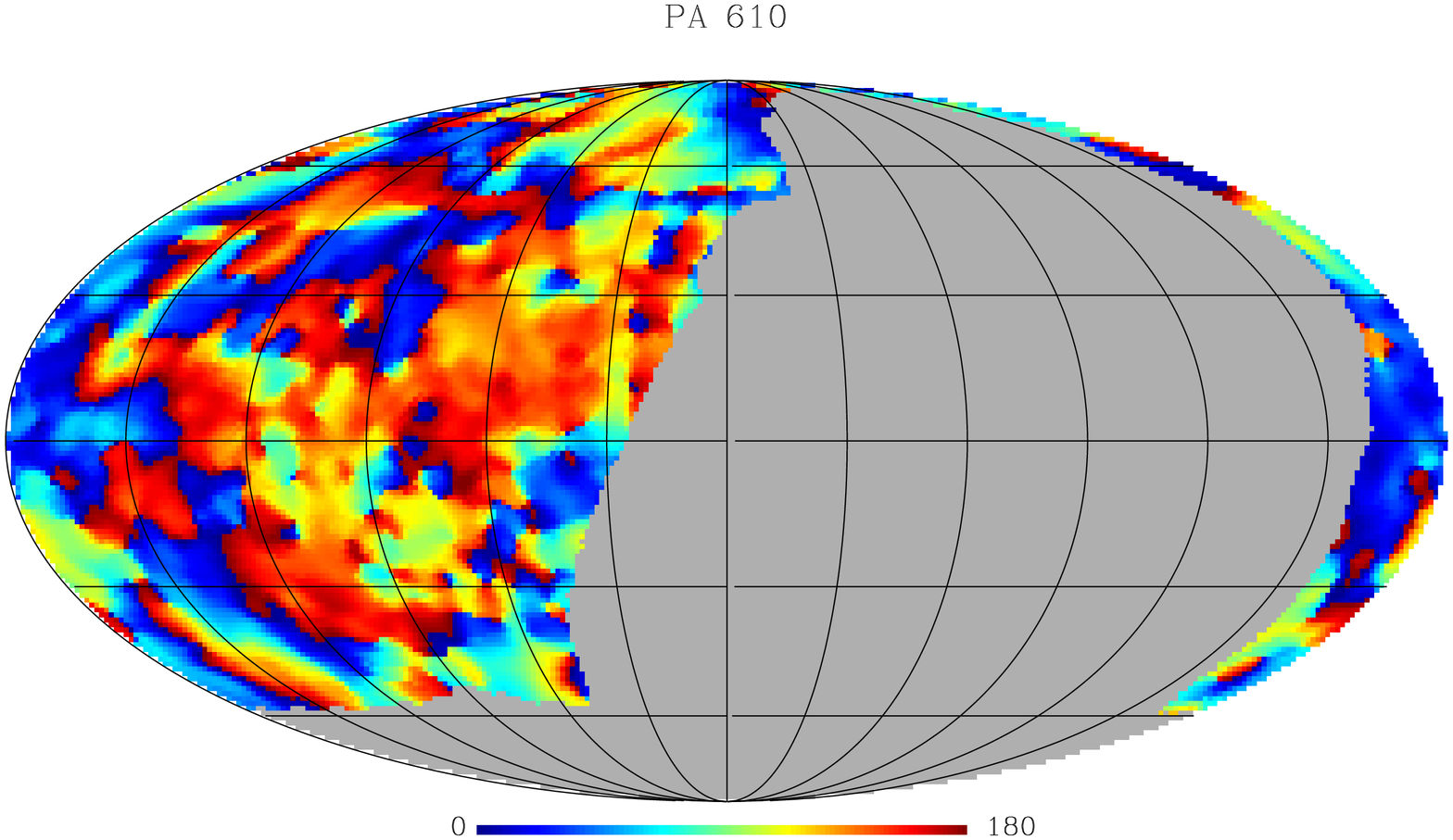}
  \includegraphics[angle=90, width=0.4\hsize]{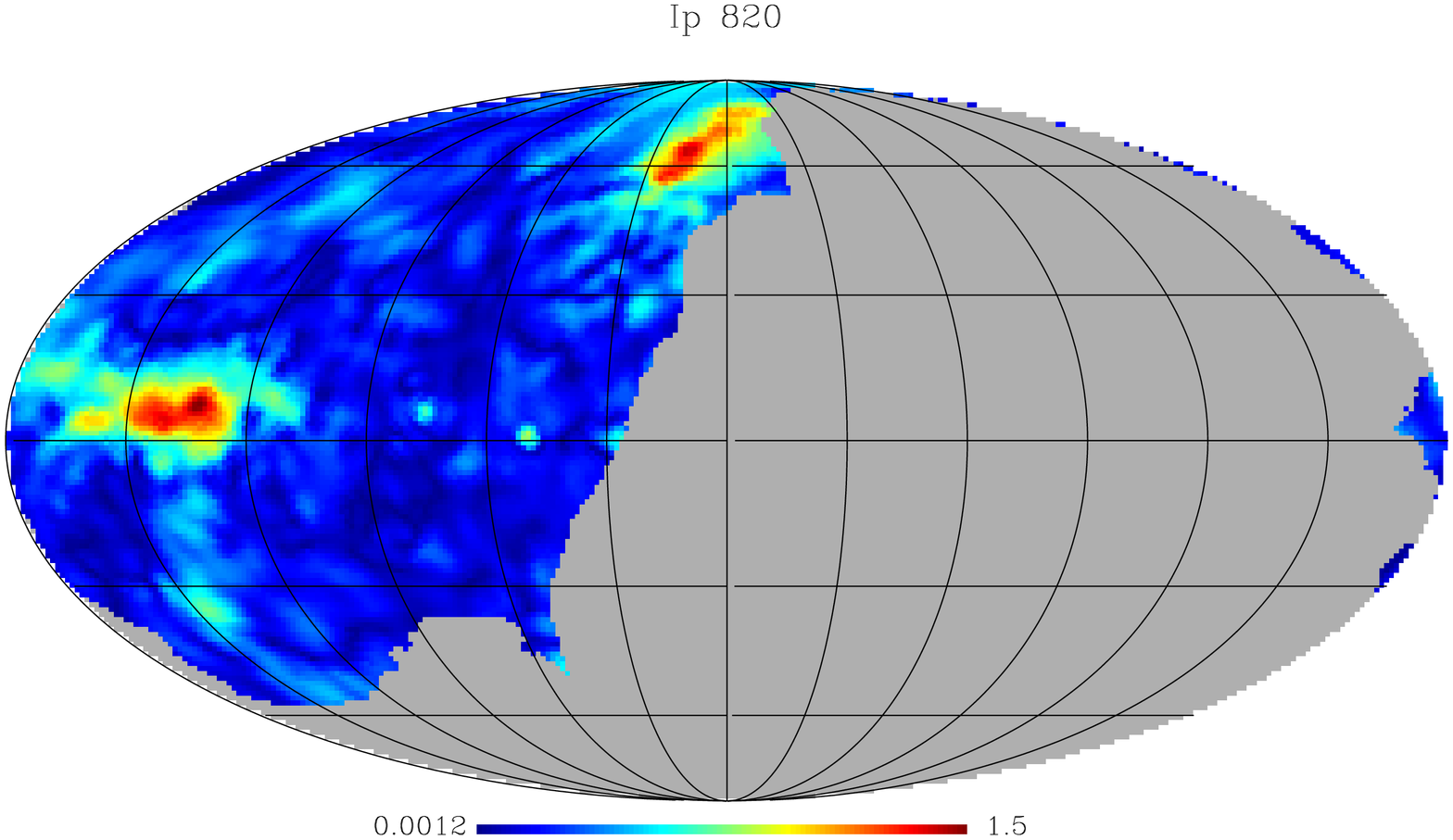}\hskip 1.0cm
  \includegraphics[angle=90, width=0.4\hsize]{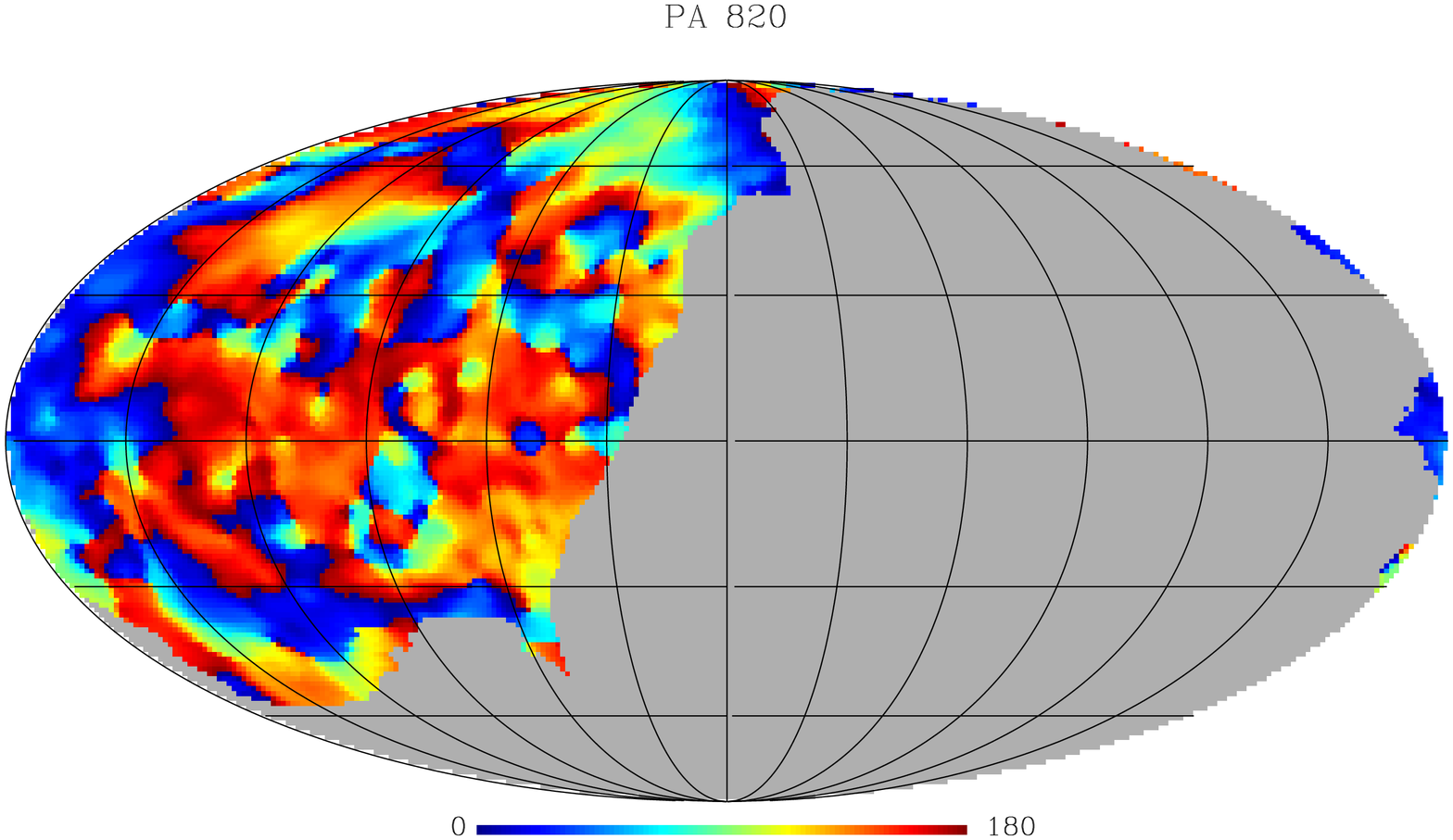}
  \includegraphics[angle=90, width=0.4\hsize]{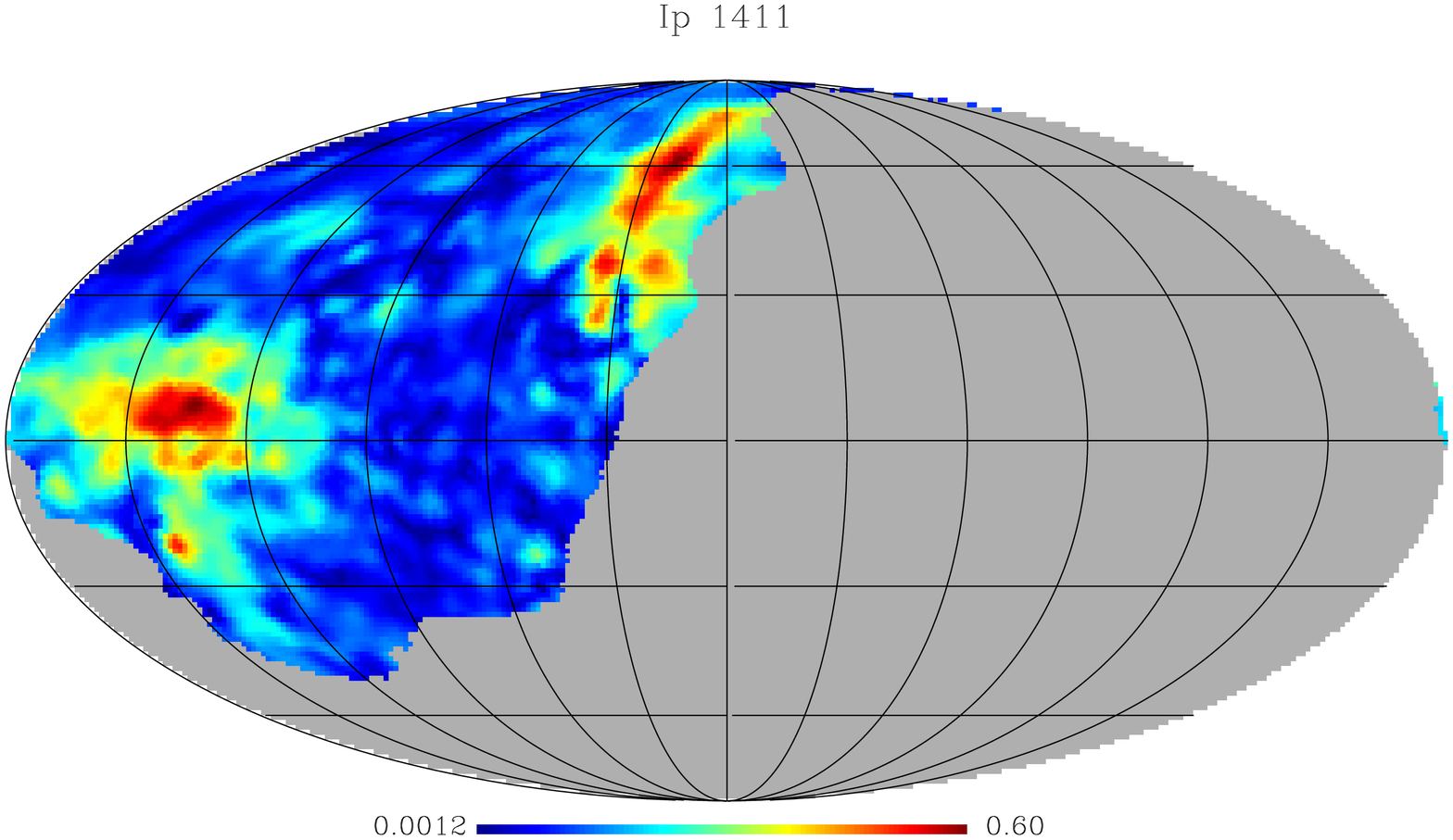}\hskip 1.0cm
  \includegraphics[angle=90, width=0.4\hsize]{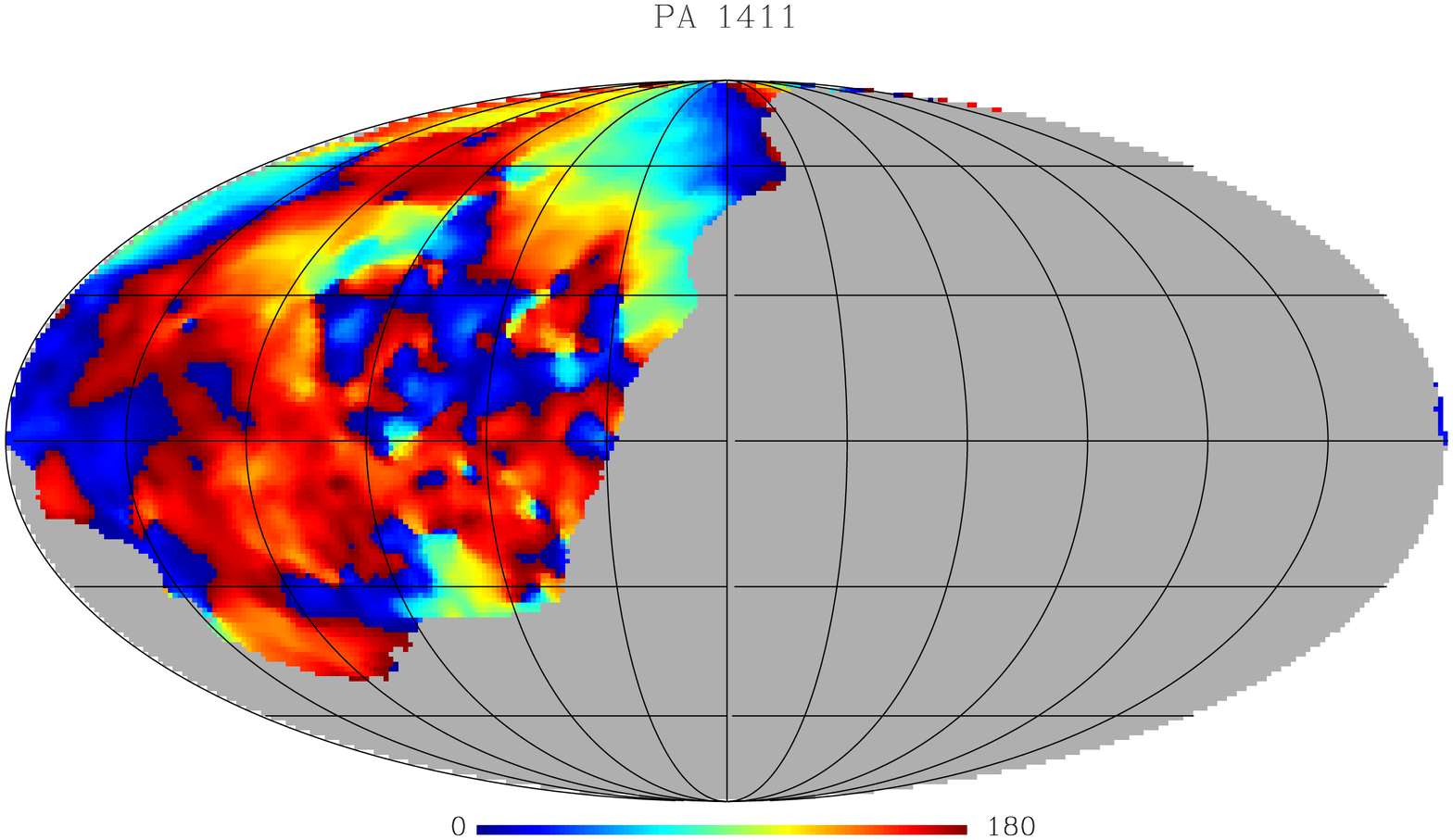}
  \caption{Polarized intensity $I^p$ (left) and polarization angle maps (right)
  formed by interpolating the \citet{bs76} data. The maps correspond to
  408, 465, 610, 820, 1411~MHz (from top to bottom). The units are Kelvin
  and degrees, respectively.
  The maps, in Galactic coordinates centred on the Galactic centre,
  have been convolved with a $4^\circ$ FWHM Gaussian filter.}
  \label{bsPIFig}
\end{figure*}

The resultant maps at the 5 frequencies are shown in
Figure~\ref{bsPIFig}. A feature is clearly visible in the Fan region at Galactic
longitude $l\sim 150^\circ$  at all frequencies.
This region is close to the area
where the line of sight is nearly perpendicular to the local Galactic
magnetic field. Hence, there is only a  small parallel magnetic field
component, and so small Faraday rotation effects.
Consequently the presence of a large non-depolarized region is not surprising,
even at the lowest frequency.

For the current study, the area of most interest is
at very high Galactic latitudes.
The North Galactic Spur, while depolarized at 408~MHz, becomes evident
with increasing frequency. Moreover, this
structure is apparent at the highest latitudes first and propagates
toward lower ones as frequency increases:
while at 610~MHz the large scale polarized structure is evident
only close to the NGP, at 820~MHz it is present
down to $b\sim 60^\circ$ and reaches $b\sim 40^\circ$
at 1411~MHz.

The polarization angle maps exhibit similar behavior.
While complex at low frequency, the
polarization angle pattern becomes more regular in the NGP region at
the highest frequencies. At 610~MHz, the
region of ordered behavior is limited to the very high latitude areas. Order
expands to lower latitudes with
increasing frequency, reaching	$b\sim 40^\circ$--$50^\circ$ at 1411~MHz.

It is easy to interpret this
in light of the previous discussion.
The randomization of the polarization angles due to Faraday rotation,
transferring the power from large to small scales,
destroys the polarized emission on the largest scales
at the lowest frequencies.
At higher frequencies, the effects of Faraday rotation decrease,
leading to the re-appearing of the large
scale structures. This starts in the areas with smaller $R\!M$.
Considering the $R\!M$ behavior with Galactic latitudes, we expect that this
would start at the very high latitudes and expand
to the mid latitudes with increasing frequency,
producing larger {\it ordered} regions 
showing the intrinsic structure.

This analysis of the \citet{bs76} data supports 
that our patch, 
given the observing parameters (at 1.4~GHz and $|b| \sim 40^\circ$), 
is in an intermediate state between
strong and negligible influence of the Faraday rotation.

\section{Implications for CMBP}\label{cmbpSec}

The $E$- and $B$-mode spectra of Section~\ref{APSSec} are
the first measurements for a high Galactic latitude patch at 1.4 GHz
on sub-degree scales (i.e. scales on which CMBP has most of the power).

Other spectra out of the Galactic plane
($b < 30^\circ$) have been measured by~\citet{have03}, 
but at lower frequency ($\sim 350$~MHz).
Their results show slopes with a very large distribution, suggesting
significant changes of Faraday rotation.
However, a decrease in slope toward higher Galactic latitudes seems to exist,
in agreement with the interpretation given here.

Our data are at a significantly higher frequency and,
being less affected by Faraday rotation features, are to date
the most suitable for extrapolation to the frequency
range of CMB measurements.

\begin{figure}
  \includegraphics[angle=0, width=1.0\hsize]{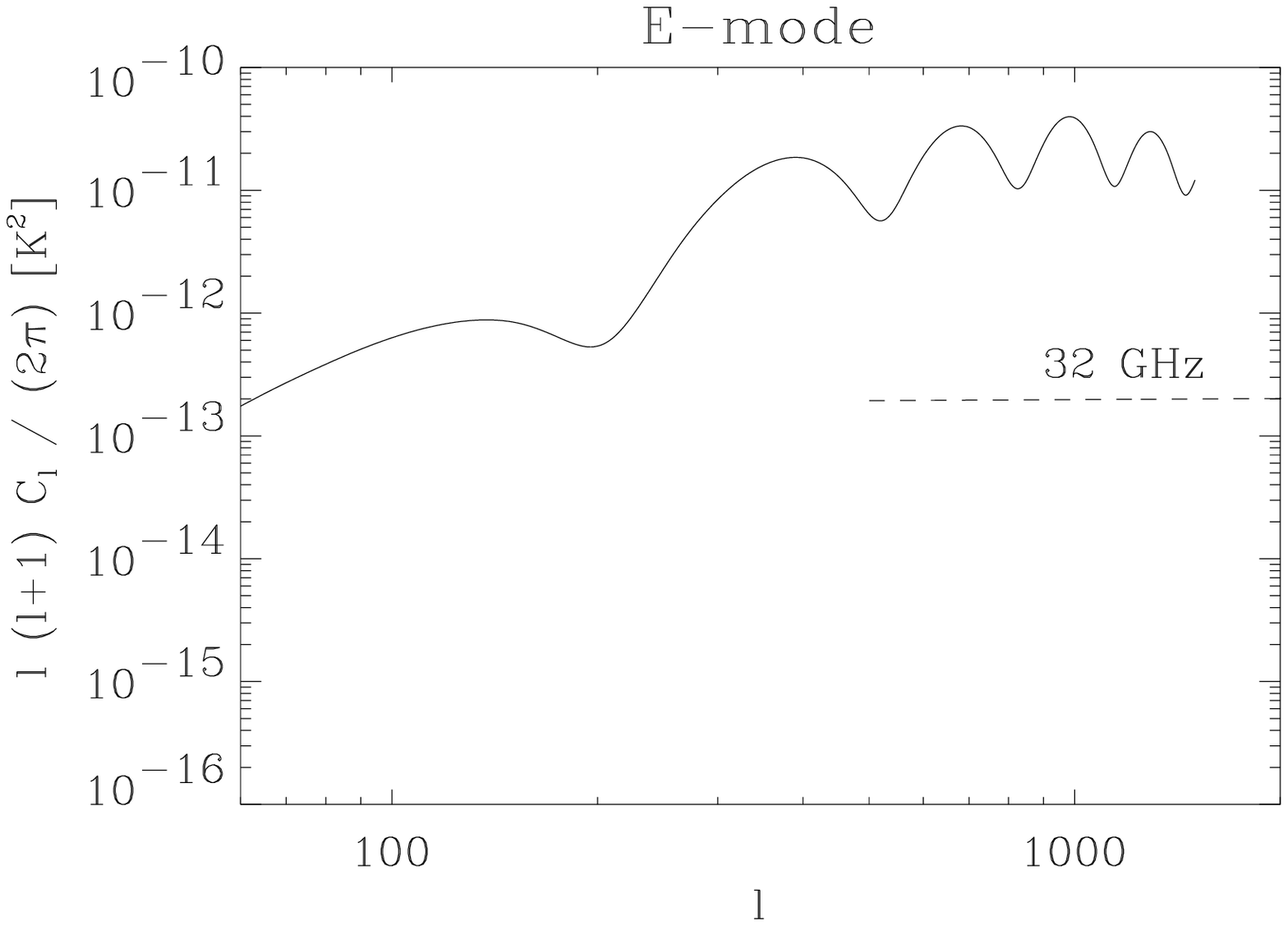}
  \includegraphics[angle=0, width=1.0\hsize]{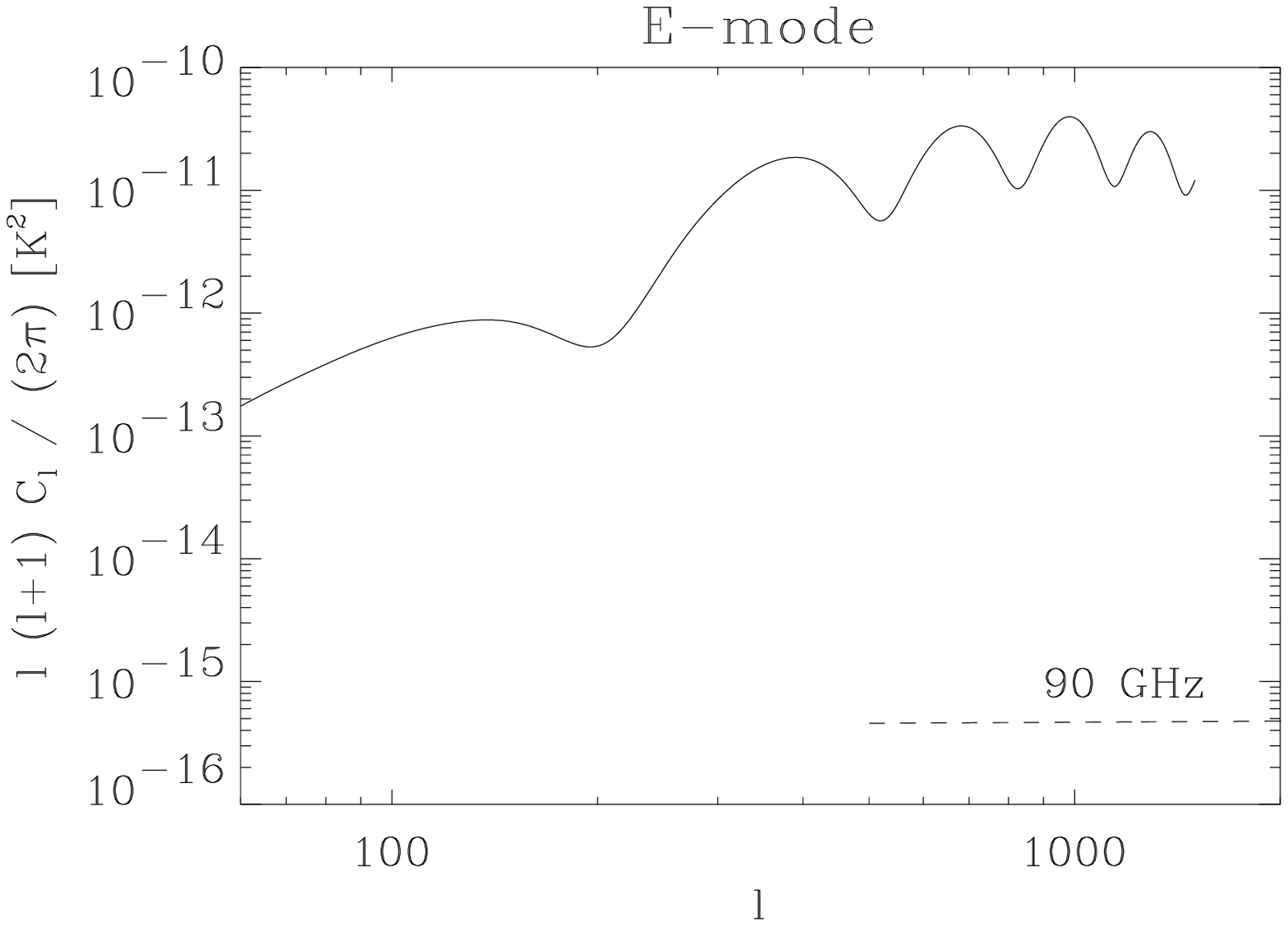}
  \caption{Fit of the $E$-mode spectrum measured for our data scaled up
	   to 32 (top) and 90~GHz (bottom).
	   The spectrum expected for CMBP with cosmological
	   parameters as from WMAP results \citep{spergel03} is also shown.}
  \label{cmbSpecFig}
\end{figure}

Figure~\ref{cmbSpecFig} shows the $E$-mode spectrum
extrapolated up to the 32 and 90~GHz frequencies
of the BaR-SPOrt experiment, assumed the synchrotron follows
a power law
\begin{equation}
   T_{\rm synch} \propto \nu^{\gamma}
\end{equation}
with spectral index $\gamma = -3.1$, typical of the 1.4--23~GHz
spectral range~\citep{bernardi04b}.
These spectra include the correction for the square of the
conversion factor
\begin{equation}
   c = \left[{2\,\sinh(x/2) \over x}\right]^2, \hskip 1cm x = h\nu / kT_{\rm cmb}
\end{equation}
transforming antenna into thermodynamic temperature of CMB
($\nu$ is the frequency and $T_{\rm cmb} = 2.726$~K).

The presence of Faraday rotation effects makes the
power in our images an upper limit to the intrinsic power on CMBP scales.
Hence, apart from errors in the frequency spectral index,
these extrapolations represent an upper limit of the contamination
on the cosmic signal.

These results suggest that the synchrotron emission should only
marginally contaminate the cosmological signal at 32~GHz, making this
patch a good target for CMBP investigations.
Even assuming an error of $\Delta\gamma = 0.2$ on the power law,
the contamination would be only a factor 2 stronger, which does not change
the conclusion.

The situation at 90~GHz is clearer still: the extrapolated spectrum is
more than 4 orders of magnitude lower than the cosmic signal, leaving
the CMBP practically uncontaminated by synchrotron pollution.
Including the steepening of the synchrotron spectral index above 23~GHz
observed by the WMAP team~\citep{bennett03b}, the
conclusion is very robust.

Such low emission of the Galactic synchtron makes this area promising
even for the weak $B$-mode.
\begin{figure}
  \includegraphics[angle=0, width=1.0\hsize]{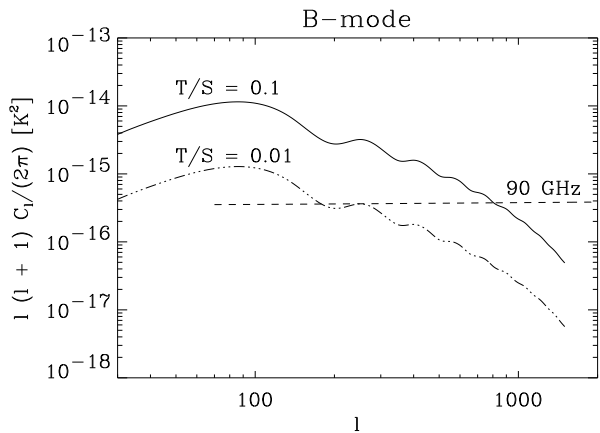}
  \caption{Fit of the $B$-mode spectrum measured for our data scaled up
	   to 90~GHz.
	   $B$-mode spectra expected for CMBP with 
	   tensor-to-scalar perturbations power ratios $T/S=0.1$ and $T/S=0.01$ 
	   are also shown. The other
	   cosmological parameters are as from WMAP results~\citep{spergel03}.}
  \label{cmbSpecBFig}
\end{figure}
Its emission has a peak near $\ell = 100$, whereas
our data cover only the $\ell = 800$--$2800$ range. To compare
 the emissions, in
Figure~\ref{cmbSpecBFig} we extrapolate
the synchrotron spectrum down to $\ell=100$ with the same slope of
Table~\ref{powFitTab}. In fact, since the slope we have measured at
1.4~GHz could be steeper than the intrinsic one,
our estimate is somewhat conservative.

Figure~\ref{cmbSpecBFig} indicates that a model
with tensor-to-scalar perturbation power ratio $T/S = 0.1$ should be well accessible
in this area, the synchrotron contribution being a factor 30 fainter than
the CMBP spectrum. This gap makes the result quite robust to errors
in the extrapolation.

Even more interesting is that
a model with $T/S = 0.01$ could be accessible in this part of the sky.
At this low level it is likely
that the leading contaminant will be the thermal dust,
whose study requires measurements at higher
frequencies (hundreds of GHz).

\section{Summary and Conclusions}\label{concSec}

We have analysed the observation by~\citet{be03} of the polarized emission at
1.4~GHz in the high Galactic
latitude area ($b\sim -40^\circ$). This represents the first
detection of the Galactic synchrotron
polarized emission carried out in a low emission area at 1.4~GHz
on this angular scale, giving us
good data to estimate
the contamination of the CMBP signal by the Galactic synchrotron
radiation.

The contamination has been evaluated through the polarized
angular power spectra $C^E$ and $C^B$. These follows a power law behavior
with spectral indeces 
$\beta_E = -1.97 \pm 0.08$ and  $\beta_E = -1.98 \pm 0.07$. 
The emission level
is about 25 times fainter than in Galactic plane regions.

Extrapolations to the CMB frequency window (30--100~GHz) gives
encouraging results:
the $E$-mode of CMBP is expected to be safely accessible
already at 32~GHz, while at 90~GHz
the margin is much larger ensuring clean measurements of CMBP.
The low contamination level makes this patch a good candidate
for the detection of the weaker $B$-mode:
the tensorial signal appears accessible for models with
$T/S > 0.01$.

The analysis of the Faraday rotation effects performed in this paper
reinforces the robustness of our estimates. The low $R\!M$ values 
measured in the patch
(typically $20$~${\rm rad}\,{\rm m}^{-2}$) do not generate
significant bandwidth depolarization. Additionally, we have carried out
a quantitative analysis of the effects caused by the randomization
of the polarization angles as a result of Faraday screens.
We find an enhancement of the power on angular scales smaller than
that of randomization, which ensures that the power we measure in
the observed patch is an upper limit of the intrinsic emission -- at
least on the scales relevant for CMB purposes.
Moreover, the analysis of the Faraday screen effects
provides a second important result:
a steepening of the power law index on angular scales smaller than
that of randomization.
The comparison between the slopes we have measured with those obtained by
other surveys suggests the patch is in an intermediate state
between significant and negligible Faraday rotation effects.

A similar conclusion is reached from an analysis of the data of~\citet{bs76}.
Depolarization effects are seen to decrease
on the large scale polarized emission
as the frequency increases from 408~MHz up to
1411~MHz. Starting from the NGP at the lowest frequencies,
the large scale emission appears down to $b\sim 40^\circ$
at 1411~MHz. This further supports the conclusion
that the Galactic latitude of the observed patch is in an intermediate
state between
significant and negligible Faraday rotation effects at 1.4~GHz.

It appears that the patch is still affected by Faraday effects, but
that it is very close to showing the intrinsic emission.
Therefore, the estimates we have given here
are conservative upper limits. Observations at higher frequency should allow
measurements which are not affected by Faraday rotation.

\section*{Acknowledgments}

This work has been carried out as part of the SPOrt experiment,
a programme funded by ASI (Italian Space Agency).
G.B. acknowledge an ASI grant. We thanks T.A.Th.~Spoelstra for
providing us the Leiden survey data and M.~Tucci for
providing us his code for angular power spectrum computation.
Part of this work is based on observations taken with
the Australia Telescope Compact Array.
The Australia Telescope Compact Array is part of the Australia Telescope,
which is funded by the Commonwealth of Australia for operation
as a National Facility managed by CSIRO.
This research has made use of the NASA/IPAC Extragalactic
Database (NED) which is operated by the Jet Propulsion Laboratory,
California Institute of Technology, under contract with the
National Aeronautics and Space Administration.
We acknowledge the use of the HEALPix and CMBFAST packages.

\bsp

\label{lastpage}


\begin{thebibliography}{99}
\bibitem[\protect\citeauthoryear{Baker \& Wilkinson}{1974}]{baker74} Baker J.R.,
	Wilkinson A., 1974, MNRAS, 167, 581
\bibitem[\protect\citeauthoryear{Bennett et al.}{2003b}]{bennett03b} Bennett C.L.,
	et al., 2003, ApJS, 148, 97
\bibitem[\protect\citeauthoryear{Bernardi}{2004}]{bernardi04} Bernardi G.,
	2004, PhD thesis, Universit\'a di Bologna
\bibitem[\protect\citeauthoryear{Bernardi et al.}{2003}]{be03} Bernardi G.,
	Carretti E., Cortiglioni S., Sault R.J., Kesteven M.J., Poppi S.,
	2003, ApJ, 594, L5
\bibitem[\protect\citeauthoryear{Bernardi et al.}{2004}]{bernardi04b} Bernardi G.,
	Carretti E., Fabbri R., Sbarra C., Cortiglioni S., Poppi S., Jonas J.L.,
	2004, MNRAS, 351, 436
\bibitem[\protect\citeauthoryear{Broten, MacLeod \& Vall\'ee}{Broten et al.}{1988}]
	{broten88} Broten N.W., MacLeod J.M., Vall\'ee, 1988,
	Ap\&SS, 141, 303
\bibitem[\protect\citeauthoryear{Brouw \& Spoelstra}{1976}]{bs76} Brouw W.N.,
	Spoelstra T.A.Th., 1976, A\&AS, 26, 129
\bibitem[\protect\citeauthoryear{Bruscoli et al.}{2002}]{br02} Bruscoli M.,
	Tucci M., Natale V., Carretti E., Fabbri R., Sbarra C., Cortiglioni S.,
	2002, New Astron., 7, 171
\bibitem[\protect\citeauthoryear{Carretti et al.}{2002}]{carretti02} Carretti E.,
	et al., 2002,
	in De Petris M. \& Gervasi M., eds,
	Experimental Cosmology at Millimeter Wavelengths, AIP Conf. Proc., 616, 140
\bibitem[\protect\citeauthoryear{Cortiglioni et al.}{2003}]{cortiglioni03} Cortiglioni S.,
	et al., 2003,
	in Warmbein~B., ed., 16th ESA Symposium on European Rocket and Balloon Programmes
	and Related Research, ESA Proc. SP-530, p. 271
\bibitem[\protect\citeauthoryear{Duncan et al.}{1997}]{duncan97} Duncan A.R.,
	Haynes R.F., Jones K.L., Stewart R.T., 1997, MNRAS, 291, 279
\bibitem[\protect\citeauthoryear{Duncan et al.}{1999}]{duncan99} Duncan A.R.,
	Reich P., Reich W., F\"urst E., 1999, A\&A, 350, 447
\bibitem[\protect\citeauthoryear{Frater, Brooks \& Whiteoak}{Frater et al.}{1992}]
	{frater92} Frater~R.H., Brooks~J.W., Whiteoak~J.B., 1992,
	J. Electrical Electron. Eng. Australia, 12, 103
\bibitem[\protect\citeauthoryear{Gaensler et al.}{2001}]{gaensler01} Gaensler B.M.,
	Dickey J.M., McClure-Griffiths N.M., Green A.J., Wieringa M.H.,
	Haynes R.F., 2001, ApJ, 549, 959
\bibitem[\protect\citeauthoryear{Giardino et al.}{2002}]{gi02} Giardino G.,
	Banday A.J., G\'orski K.M., Bennett K., Jonas J.L., Tauber J.,
	2002, A\&A, 387, 82
\bibitem[\protect\citeauthoryear{G\'orski, Hivon \& Wandelt}{G\'orski et al.}{1999}]{go99}
	G\'orski K.M.,
	Hivon E., Wandelt B.D., 1999,
	in Banday~A.J., Sheth~R.S. \& da~Costa~L., eds, Proceedings of the MPA/ESO Cosmology
	Conference Evolution of Large-Scale Structure, Print Partners Ipskamp,
	NL, p.~37, astro-ph/9812350
\bibitem[\protect\citeauthoryear{Griffith \& Wright}{1993}]{griffith93} Griffith M.R.,
	Wright A.E., 1993, AJ, 105, 5
\bibitem[\protect\citeauthoryear{Han et al.}{1997}]{han97} Han J.L., Manchester R.N.,
	Berkhuijsen E.M., Beck R., 1997, A\&A, 322, 98
\bibitem[\protect\citeauthoryear{Han}{2004}]{han04} Han J.L., 2004,
	in Uyan{\i}ker~B., Reich~W. \& Wielebinski~R., eds,
	The Magnetized Interstellar Medium, Copernicus GmbH, p. 3
\bibitem[\protect\citeauthoryear{Haslam et al.}{1982}]{haslam82} Haslam C.G.T.,
	Stoffel~H., Salter~C.J., Wilson~W.E., 1982, A\&AS, 47, 1
\bibitem[\protect\citeauthoryear{Haverkorn, Katgert \& de Bruyn}{Haverkorn et al.}{2003}]{have03}
	Haverkorn M., Katgert P., de Bruyn A.G., 2003, A\&A, 403, 1045
\bibitem[\protect\citeauthoryear{Hockney \& Eastwood}{1981}]{he81} Hockney R.W.,
	Eastwood J.W., 1988, Computer Simulation Using Particles
	(New York: McGraw-Hill)
\bibitem[\protect\citeauthoryear{Huchra et al.}{2003}]{huchra03} Huchra~J., Jarrett~T.,
	Skrustikie~M., Cutri~R., Schneider~S., Macri~L., Steining~R., Mader~J.,
	2003, in IAUS Symposium 216, ASP Conf. Series, S-216, p.~172
	The Magnetized Interstellar Medium, Copernicus GmbH, p. 13
\bibitem[\protect\citeauthoryear{Johnston-Hollitt, Hollitt \& Ekers}
				{Johnston-Hollitt et al.}{2004}]{john04} Johnston-Hollitt M.,
	Hollitt C.P., Ekers R.D., 2004,
	in Uyan{\i}ker~B., Reich~W. \& Wielebinski~R., eds,
	The Magnetized Interstellar Medium, Copernicus GmbH, p. 13
\bibitem[\protect\citeauthoryear{Jonas, Baart \& Nicolson}{Jonas et al.}{1998}]{jonas98}
	Jonas J.L., Baart E.E., Nicolson G.D., 1998, MNRAS, 297, 977
\bibitem[\protect\citeauthoryear{Maddox et al.}{1990}]{maddox90} Maddox S.J.,
	Sutherland W.J., Efstathiou G., Loveday J., 1990, MNRAS, 243, 692
\bibitem[\protect\citeauthoryear{Masi et al.}{2002}]{masi02}
	Masi S., et al., 2002,
	in Cecchini~S., Cortiglioni~S., Sault~R.J., Sbarra~C., eds,
	Astrophysical Polarized Backgrounds, AIP Conf. Proc., 609, 122
\bibitem[\protect\citeauthoryear{Reich \& Reich}{1986}]{reich86}
	Reich P., Reich W., 1986, A\&AS, 63, 205
\bibitem[\protect\citeauthoryear{Reich, Testori \& Reich}{Reich et al.}{2001}]{reich01}
	Reich P., Testori~J.C., Reich~W., 2001, A\&A, 376, 861
\bibitem[\protect\citeauthoryear{Reich et al.}{2004}]{reich04}
	Reich W., F\"urst E., Reich P., Uyan{\i}ker B.,
	Wielebinski R., Wolleben M., 2004,
	in Uyan{\i}ker~B., Reich~W. \& Wielebinski~R., eds,
	The Magnetized Interstellar Medium, Copernicus GmbH, p.~45
\bibitem[\protect\citeauthoryear{Sault, Teuben \& Wright}{Sault et al.}{1995}]{sault95}
	Sault~R.J., Teuben~P.J., Wright~M.H.C., 1995,
	in Shaw~R., Payne~H.E., Hayes~J.J.E., eds,
	ASP Conf. Series, 77, p. 433
\bibitem[\protect\citeauthoryear{Seljak}{1997}]{seljak97} Seljak~U.,
	1997, ApJ, 482, 6
\bibitem[\protect\citeauthoryear{Spergel et al.}{2003}]{spergel03} Spergel D.N.,
	et al., 2003, ApJS, 148, 175
\bibitem[\protect\citeauthoryear{Sun \& Han}{2004}]{sun04} Sun~X.H., Han~J.L., 2004,
	in Uyan{\i}ker~B., Reich~W. \& Wielebinski~R., eds,
	The Magnetized Interstellar Medium, Copernicus GmbH, p. 25
\bibitem[\protect\citeauthoryear{Taylor et al.}{2003}]{taylor03} Taylor A.R.,
	 et al., 2003, AJ, 125, 3145
\bibitem[\protect\citeauthoryear{Tegmark et al.}{2000}]{tegmark00} Tegmark~M.,
	Eisenstein~D.J., Hu~W., de Oliveira-Costa~A., 2000, ApJ, 530, 133
\bibitem[\protect\citeauthoryear{Testori, Reich \& Reich}{Testori et al.}{2004}]
	{testori04} Testori J.C., Reich P., Reich W., 2004,
	in Uyan{\i}ker~B., Reich~W. \& Wielebinski~R., eds,
	The Magnetized Interstellar Medium, Copernicus GmbH, p.~57
\bibitem[\protect\citeauthoryear{Tucci et al.}{2002}]{tu02} Tucci M.,
	Carretti E., Cecchini S., Nicastro L., Fabbri R., Gaensler B.M.,
	Dickey J.M., McClure-Griffiths N.M., 2002, ApJ, 579, 607
\bibitem[\protect\citeauthoryear{Uyan{\i}ker et al.}{1999}]{uyaniker99} Uyaniker B.,
	F\"urst E., Reich W., Reich P., Wielebinski R., 1999,
	A\&AS, 138, 31
\bibitem[\protect\citeauthoryear{Wieringa et al.}{1993}]{wieringa93} Wieringa~M.H.,
	 de~Bruyn~A.G., Jansen~D., Brouw~W.N., Katgert~P., 1993, A\&A, 268, 215
\bibitem[\protect\citeauthoryear{Wolleben et al.}{2004}]{wolleben04}
	Wolleben M., Landecker T.L., Reich W., Wielebinski R., 2004,
	in Uyan{\i}ker~B., Reich~W. \& Wielebinski~R., eds,
	The Magnetized Interstellar Medium, Copernicus GmbH, p.~51
\bibitem[\protect\citeauthoryear{Zaldarriaga}{1998}]{zaldarriaga98} Zaldarriaga M.,
	PhD thesis, MIT
\bibitem[\protect\citeauthoryear{Zaldarriaga \& Seljak}{1997}]{zaldarriaga97} Zaldarriaga M.,
	Seljak~U., 1997, PRD, 55, 1822
\bibitem[\protect\citeauthoryear{Zaldarriaga, Spergel \& Seljak}{Zaldarriaga et al.}{1997}]
	{zaldarriaga97b} Zaldarriaga~M., Spergel~D.N., Seljak~U., 1997, ApJ, 488, 1
\end{thebibliography}
\end{document}